\documentclass{jkas}
\usepackage{xcolor}
\usepackage{chngcntr}

\def\year{2020} 
\def\month{April} 
\def\volume{00} 
\def\issue{0} 
\def\beginpage{1} 
\def\endpage{99} 
\def\received{---} 
\def\accepted{---} 
\date{Received \received; accepted \accepted}

\newcommand\ha{H$\alpha$}
\newcommand\hb{H$\beta$} 
\newcommand\hst{{\it HST}}
\def\lum{erg s$^{-1}$}
\def\farcs{\hbox{$.\mkern-4mu^{\prime\prime}$}}

\def\ser{S\'{e}rsic}
\def\mlr{$M_{\rm BH}$--$M_{I,\rm bul}$}
\def\mlrr{$M_{\rm BH}$--$M_{R,\rm bul}$}
\def\mlb{$M_{\rm BH}$--$M_{\rm bul}$}

\def\mmr{$M_{\rm BH}$--$M_{*,\rm bul}$}
\def\mue{$\langle\mu_{\rm e}\rangle$}
\def\re{$R_{\rm e}$}
\newcommand\ct[1]{\multicolumn{1}{c}{#1}}

\title{
Relation between Black Hole Mass and Bulge Luminosity in Hard X-ray selected Type 1 AGNs
}

\author[1]{Suyeon Son}
\author[1]{Minjin Kim}
\author[2]{Aaron J. Barth}
\author[3,4]{Luis C. Ho}

\affil[1]{Department of Astronomy and Atmospheric Sciences, 
College of Natural 
Sciences, Kyungpook National University, Daegu 41566, Korea; 
\email{mkim.astro@gmail.com}}
\affil[2]{Department of Physics and Astronomy, University of California, 
Irvine, CA 92697-4575, USA}
\affil[3]{Kavli Institute for Astronomy and Astrophysics, Peking University, 
Beijing 100871, China}
\affil[4]{Department of Astronomy, School of Physics, Peking University, Beijing 
100871, China}

\def\corrauthor{
M. Kim
}

\def\runningauthor{
Author Son et al.
}

\def\runningtitle{
\mlr\ relation of type 1 AGNs
}

\def\keywords{
black hole physics --- 
galaxies: active --- galaxies: Seyfert --- quasars
}

\def\abstracttext{
Using $I$-band images of 35 nearby ($z<0.1$) type 1 active galactic nuclei 
(AGNs) obtained with {\it Hubble Space Telescope}, selected from the 
70-month Swift-BAT X-ray source catalog, we investigate the photometric 
properties of the host galaxies. With a careful treatment of the point-spread 
function (PSF) model and imaging decomposition, we robustly measure the 
$I$-band brightness and the effective radius of bulges in our sample. 
Along with black hole (BH) mass estimates from single-epoch spectroscopic 
data, we present the relation between BH mass and $I$-band bulge luminosity 
(\mlr\ relation) of our sample AGNs. We find that our sample lies 
offset from the \mlr\ relation of inactive galaxies by 0.4 dex, i.e., 
at a given bulge luminosity, the BH mass of our sample is systematically
smaller than that of inactive galaxies.
We also demonstrate that the zero point offset in the \mlr\ relation with 
respect to inactive galaxies is correlated with the Eddington ratio. 
Based on the Kormendy relation, we find that the mean 
surface brightness of ellipticals and classical bulges in our sample is 
comparable to that of normal galaxies, revealing that bulge brightness is 
not enhanced in our sample. As a result, we conclude that the deviation in 
the \mlr\ relation from inactive galaxies is possibly because the scaling 
factor in the virial BH mass estimator depends on the Eddington ratio. 
}

\begin{document}
\jkashead 


\section{Introduction}
Supermassive black holes (SMBHs) are 
ubiquitous at the centers of massive galaxies, and their mass is tightly 
correlated with various physical properties of bulges, such as luminosity, 
stellar mass, and velocity dispersion (\citealt{kormendy_2013}). While these 
scaling relations imply a physical connection between SMBH and the host 
galaxy in terms of their formation and evolution, the physical origin of 
the scaling relations is still under debate.
Energetic feedback from an active galactic nucleus (AGN) is one of the 
favored mechanisms to drive the co-evolution of SMBH and the host galaxy 
(e.g., \citealt{dimatteo_2005}). In this light, exploring the scaling 
relations in AGN is of great importance to unveil the origin of the 
BH-host relations. 

Several scaling relations have been widely used to understand the 
causal connection between SMBHs and their host galaxies in active galaxies, thanks to the 
relative ease of estimating the BH mass of type 1 AGN, which exhibit blue continuum from the accretion disk and 
broad emission lines in the spectrum. For example, using 
the relation between BH mass and physical properties of the bulge 
(e.g., stellar velocity dispersion and bulge luminosity) from distant 
AGNs, previous studies argued that BH growth precedes galaxy growth (e.g., 
\citealt{peng_2006, woo_2008, park_2014}). Using AGN in the local universe, 
several studies reported that active galaxies deviate from inactive galaxies in the 
\mlb\ relation, revealing that either BH mass is systematically 
undermassive or bulge luminosity is overluminous compared to normal 
galaxies (e.g., \citealt{kim_2008b, kim_2019}).

On the other hand, a relation between the effective radius and 
the mean surface brightness within the effective radius (Kormendy relation;
\citealt{kormendy_1977}) has been extensively used to probe young stellar 
populations in host galaxies of nearby AGNs (e.g., \citealt{kim_2019, 
zhao_2019, zhao_2021}). For example, \citet{kim_2019} argued that bulges 
regardless of the bulge types in galaxies hosting type 1 AGNs tend to be 
brighter than normal galaxies inferred from the Kormendy relation, possibly 
due to recent star formation. 
However, using the same relation, \citet{zhao_2021} demonstrated 
that only pseudo-bulges in active galaxies are overluminous. Interestingly, 
this trend is consistent with the result for type 2 AGNs in 
\citet{zhao_2019}.  Overall, different studies reached somewhat 
different conclusions regarding the stellar population in galaxies hosting AGNs, possibly due 
to the diverse sample properties and filters ($R$ and $I$) used in the 
previous studies (e.g., \citealt{zhao_2019}).

Nevertheless, previous studies for nearby AGNs included some distant AGNs 
($z \leq 0.35-0.5$), which can still be affected by cosmic evolution
(e.g., \citealt{bennert_2010, park_2014}). Additionally, the sample was 
dominated by AGNs selected from the UV/optical survey (e.g., 
\citealt{schmidt_1983, lyke_2020}), possibly biased toward bright and less 
obscured AGNs. These factors can introduce some unknown 
biases in establishing the BH$-$host relation. In this study, we make use of 
relatively unbiased and nearby AGNs ($z<0.1$) drawn from a hard X-ray survey
(\citealt{koss_2017}). Moreover, the imaging data used in this study is 
obtained with uniform and consistent observations, which can allow us to explore the 
BH-host relation with a minimum bias due to the sample selection, 
cosmic evolution, and diverse imaging quality.

In this study, we revisit two scaling relations (BH mass$-$bulge luminosity 
and Kormendy relations) of nearby type 1 AGNs using $I$-band images
obtained with {\it Hubble Space Telescope} (\hst). In \S{2}, we present 
the physical properties of the sample, and description of the observation and 
data reduction. In \S{3}, we present the methods for the imaging 
decomposition. We present \mlr\ and Kormendy relations of our sample 
by comparing them with inactive galaxies in \S{4}. In \S{5}, we discuss 
the physical origins of systematic offsets in the scaling relations. 
A summary and conclusions are given in \S{6}.
We adopt the following cosmological parameters: $H_0=100h=67.8$ km 
${\rm s}^{-1}$ ${\rm Mpc}^{-1}$, $\Omega_m=0.308$, and $\Omega_\Lambda=0.692$ 
(\citealt{planck_2016}).

\section{Sample and Data}

\subsection{Sample}
Our sample contains nearby AGNs ($z<0.1$) drawn from the 70-month Swift-BAT 
X-ray source catalog (\citealt{koss_2017}). We utilized images obtained 
with \hst\ as a part of the gap-filler snapshot 
program (HST program 15444), in which the imaging survey of nearby hard 
X-ray selected AGNs was conducted (\citealt{kim_2021}).
Due to less severe attenuation in the hard X-ray band, 
the parent sample is thought to be relatively unbiased in terms of obscuration, 
making the Swift-BAT sample ideal to study the physical connection between 
unobscured (type 1) and obscured (type 2) AGNs. The main purpose of the \hst\ 
survey is to study the photometric properties of the host and their 
connection with the properties of AGNs. Thus, the sample for the survey 
was selected to contain nearby AGNs ($z<0.1$). Owing to its 
uniform selection and proximity, the \hst\ images of the sample are suitable 
to achieve the goal in this study, investigating the relation between the 
BH mass and bulge luminosity. 

In this light, we only included objects with spectroscopic measurements 
of broad emission lines (type 1 AGNs). Type 1.8/1.9 AGNs are excluded 
because the physical properties of their broad line regions are 
distinct from those in canonical type 1 AGNs, which can introduce 
systematic uncertainty in the BH mass measurement (e.g., 
\citealt{ho_2008}). 
Imaging data from program 15444 obtained before May 2020 are utilized in this study. 
Our final sample, originally drawn from \citet{koss_2017}, 
consists of 35 objects with a median redshift of 
$\sim0.046$ and median bolometric luminosity of $\sim 10^{44.9}$ \lum\ 
(see Tables \ref{tab:table1} and \ref{tab:table2}). The bolometric luminosity is inferred 
from the intrinsic hard X-ray luminosity (see \S{2.3}).

\subsection{Observations and Data Reduction}
The \hst\ images were taken with the Wide Field Channel (WFC) of the 
Advanced Camera for Surveys (ACS). Its field of view 
(202$^{\prime\prime}\times$202$^{\prime\prime}$) is large enough
to cover the target galaxies. To avoid contamination from the extended 
emission and maximize the light ratio of the host to the nucleus, we adopted
the F814W filter, corresponding to $I$-band. We obtained two dithered 
images with an exposure time of 337 s for efficient removal of cosmic rays
and correction for hot pixels. An additional image with an exposure of 5 s 
was acquired for each target to avoid the saturation of the bright nucleus.

The basic data reduction, which includes bias subtraction, flat-fielding, 
and correction for the charge-transfer efficiency, was performed using 
the Pyraf-based \texttt{STSDAS} package. Cosmic rays were removed from each exposure
by adopting the l.a.cosmic algorithm developed by \citet{vandokkum_2001}. 
The spatial offset between the two long exposures was computed using the \texttt{TweakReg} task, and the two exposures were combined after the distortion  
correction using \texttt{AstroDrizzle}. The pixel size in the combined 
image is 0\farcs05.
The sub-array images with short exposure time suffer heavily
from a strong pattern noise in the bias, leaving a series of stripes in the 
raw images. To solve this problem, we removed this noise using 
\texttt{ACS\_DESTRIPE\_PLUS} (\citealt{grogin_2010}). 
The saturated pixels at the AGN core in the long-exposure images 
were replaced by the corresponding pixels in the short-exposure images.
Details of the observations and data reduction are described in
\citet{kim_2021}.

\begin{figure*}[htp]
\centering
\includegraphics[width=75mm]{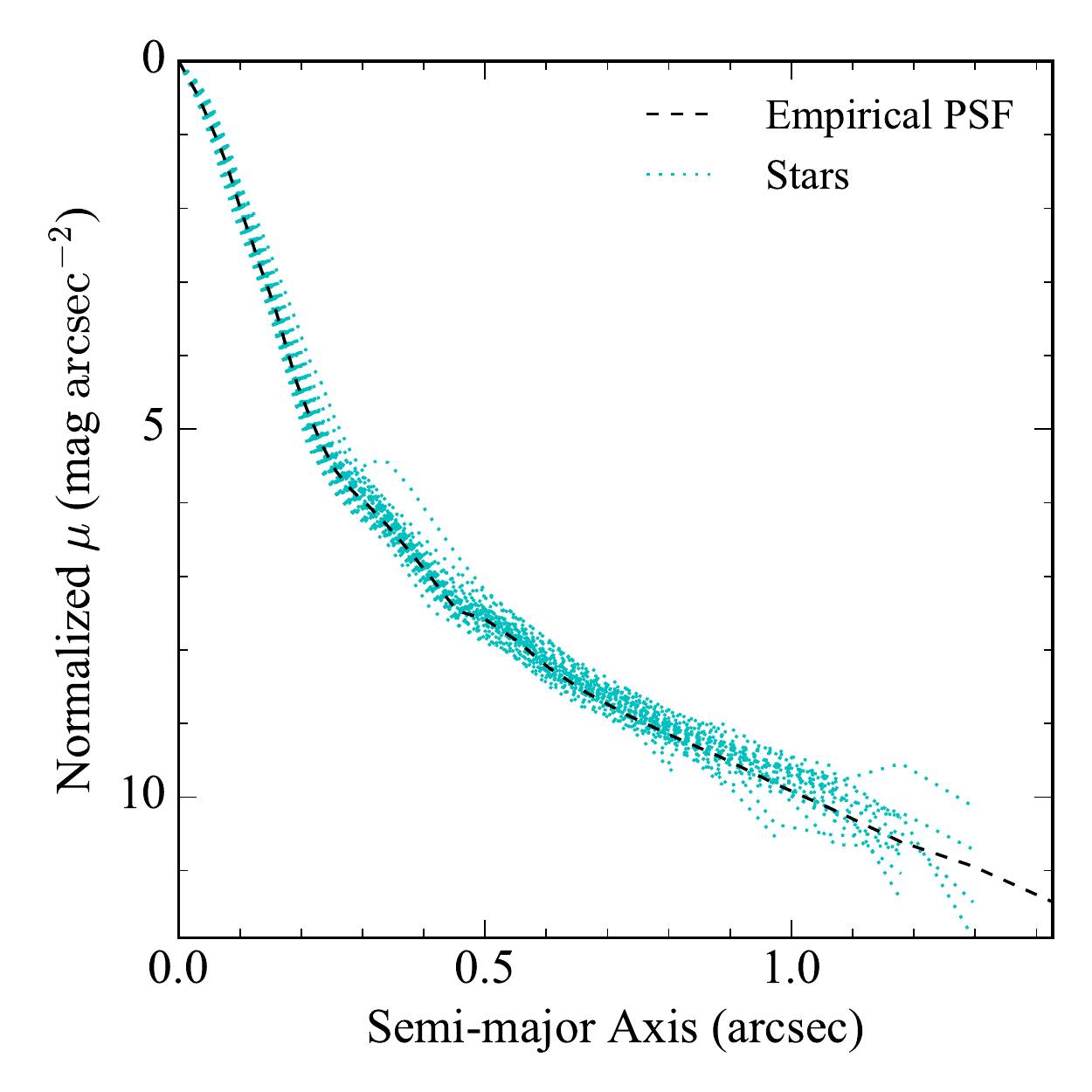}
\includegraphics[width=75mm]{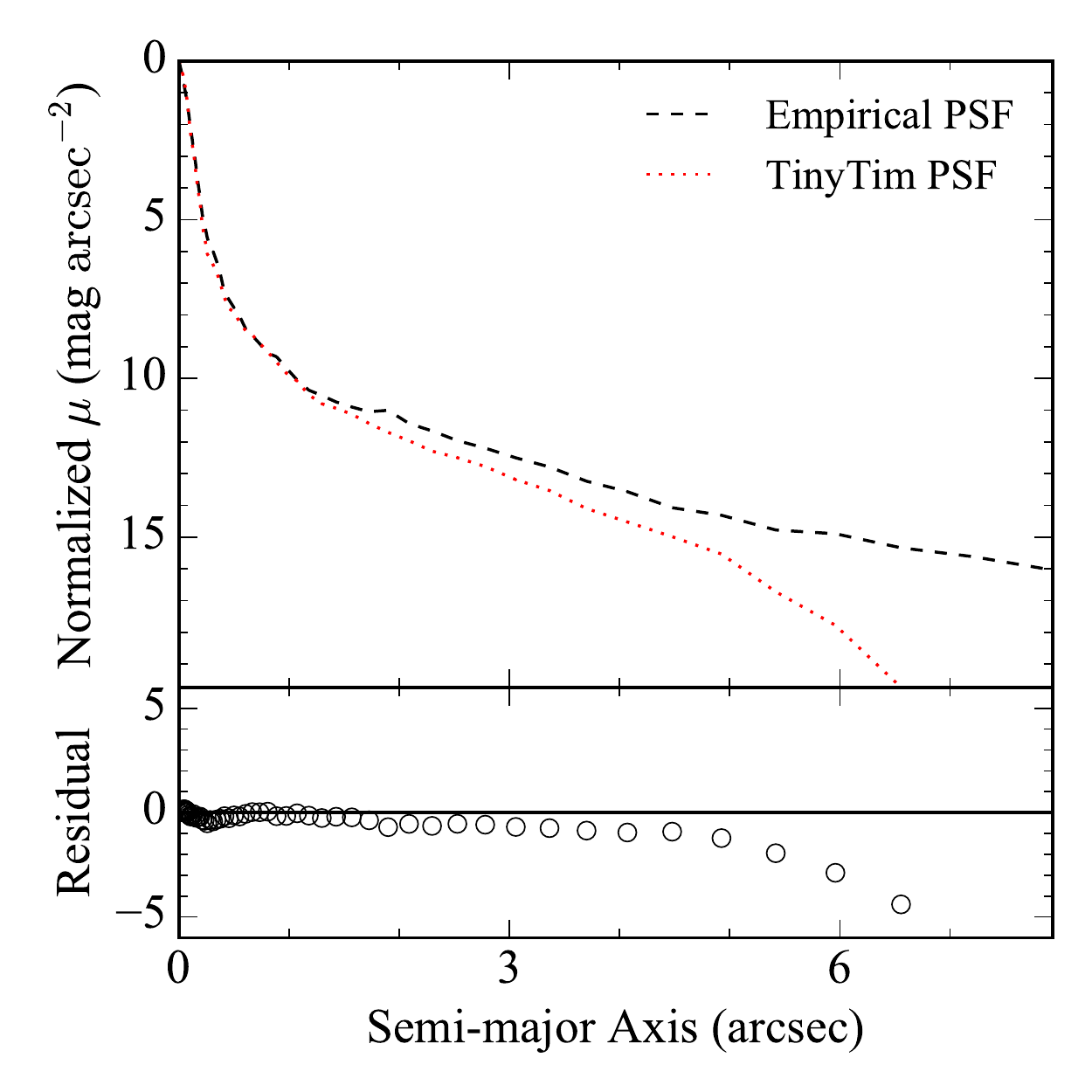}
\caption{Surface brightness profiles of various PSFs.
In the left panel, the central part of the empirical PSF and those of
individual stars are represented by a black dashed line and 
cyan dotted lines, respectively. In the top-right panel, the profiles 
of the empirical PSF and TinyTim PSF are shown in black dashed line
and red dotted line, respectively. The residuals are shown in 
the bottom-right panel. 
\label{fig:fig1}}
\end{figure*}

\subsection{BH mass and Eddington Ratio}
If the gas in the broad line region (BLR) is virialized, BH mass can be 
estimated from the combination of the velocity dispersion ($\sigma$) and 
the radius of the BLR ($R_{\rm BLR}$) as 
$M_{\rm BH} = f \sigma^2 R_{\rm BLR}/G$, 
where $f$ is a scaling factor determined by the kinematics and structure of 
the BLR. Due to the tight correlation between $R_{\rm BLR}$ and 
AGN luminosity, AGN luminosity can be a surrogate for $R_{\rm BLR}$ 
(\citealt{kaspi_2000, bentz_2013}; but see \citealt{du_2016, fonseca_2020}). 
Therefore, BH mass 
can be computed with the width and luminosity of the broad emission lines 
(\hb\ and \ha\ in this study) derived from single-epoch spectra. 
However, the scaling factor ($f$) is somewhat uncertain and has been 
empirically 
determined by assuming that, for example, active galaxies follow the same 
BH mass--stellar velocity dispersion relation of inactive galaxies. 
\citet{ho_2015} argued that 
the scaling factor ($f$) depends on the bulge types 
(i.e., classical bulge vs. pseudo-bulge). Therefore, in this study, 
we adopted two different scaling 
factors for the BH mass estimation according to their bulge types. We utilize 
the fluxes and widths of the broad \ha\ emission, estimated from the 
decomposition of the optical spectra (\citealt{koss_2017}). 
If the broad \ha\ emission is unavailable, we instead used the 
spectral measurements of broad \hb\ emission. The conversion from the fluxes 
of broad emission lines to the luminosity at 5100 \AA\ was performed using the 
conversion factors from \citet{greene_2005}. The typical uncertainty on
virial mass estimates is of $\sim0.4$ dex (e.g., \citealt{vestergaard_2006}).

Bolometric luminosities of AGNs are inferred from the hard X-ray luminosity 
estimated in the 14-195 keV band, corrected for absorption
(\citealt{ricci_2017}). For the bolometric correction, we adopted a 
single conversion factor of 8 (i.e., 
$L_{\rm bol}=8\times L_{14-195 \rm keV}$; \citealt{ricci_2017}).
The median Eddington ratio of the entire sample is 0.06.

\section{Analysis}
\subsection{PSF Generation}
To robustly estimate the bulge brightness in the images of type 1 AGNs,
a careful decomposition of the bright nuclear component, modeled by a 
PSF, is essential. The PSF can be constructed 
in two different ways: (1) an empirical PSF derived from stars obtained in 
the same observing condition (e.g., filters and detectors); 
(2) a synthetic PSF modeled by \texttt{TinyTim} software
(TinyTim PSF; \citealt{krist_2011}). The TinyTim 
PSF has been widely used as it represents the central part of the PSF 
relatively well. However, it often underestimates the surface brightness
in the halo of the PSF, which can naturally lead to the miscalculation of
underlying host brightness (e.g., \citealt{kim_2008a, zhao_2021}).  
Therefore, in this study, we employed an empirical PSF to model the 
nucleus.  

To generate the empirical PSF, we initially selected numerous sufficiently 
bright but unsaturated stars from the science images. Then,
through visual inspection, extended sources and stars with close neighbors 
are manually excluded, finally with 44 stars remaining. We use \texttt{IRAF}
to eliminate faint nearby sources around the target stars, adjust the 
scale among them, and combine the images for generating the PSF. By 
comparing the surface brightness profile of the generated PSF and those 
of individual stars obtained with \texttt{Ellipse} within \texttt{IRAF}, 
we confirmed that the profile in the central part of the PSF is 
well represented by the empirical PSF (Fig. \ref{fig:fig1}).
However, the wing of the PSF is rather dominated by noise, 
which can easily mimic the signal from the host galaxy. Therefore, 
to characterize the outer part of the PSF, we additionally utilized a 
moderately saturated star. To determine the relative scaling between the 
initial PSF and the saturated star, we modeled the 2D surface brightness 
profiles of both objects with the profile of the initial PSF using 
\texttt{GALFIT} (\citealt{peng_2002, peng_2010}). 
During the fit, the saturated pixels and charge-bleeding regions near the PSF core were properly 
masked out. Finally, we constructed the full PSF by replacing the masked
region in the image of the saturated star with the scaled initial PSF. 

To validate the empirical PSF, we compared its surface brightness
profile with that of the TinyTim PSF (Fig. \ref{fig:fig1}). Although two PSFs are
in good agreement with each other in the central region, we found that
there are substantial differences between the two PSFs in the extended wings at $r > 1^{\prime\prime}$, 
as expected. This comparison demonstrates that the empirical PSF is 
more suitable for our study. 

\begin{figure*}[htp]
\centering
\vspace{-5.0cm}
\includegraphics[width=180mm,page=1]{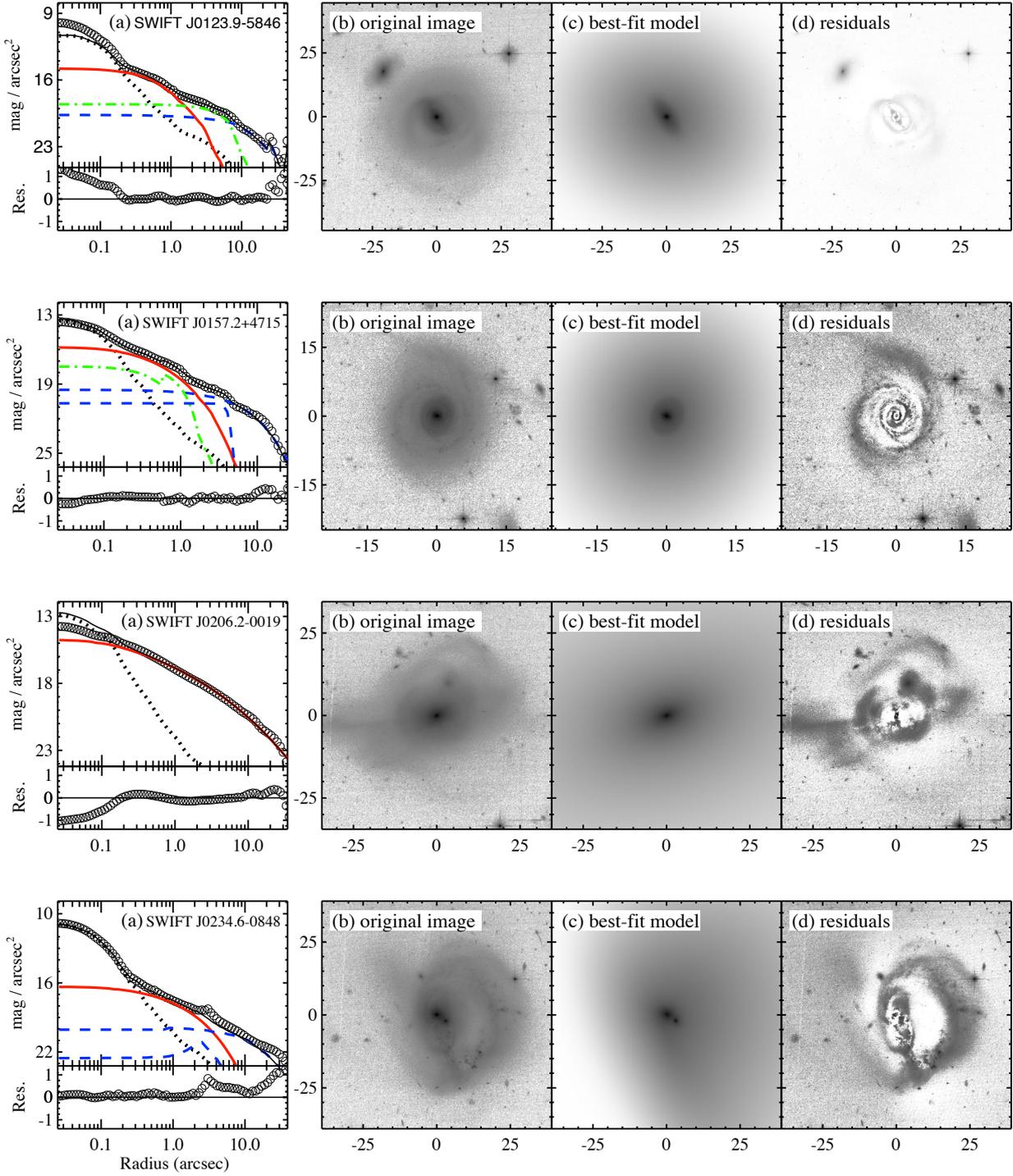}
\vspace{-1.5cm}
\caption{Results of imaging decomposition with \texttt{GALFIT}. 
(a) In the top panel, surface brightness profiles of the original image 
(open circles), nucleus (dotted line), and bulge (red line) are displayed. 
If present, bar 
and disk are denoted by green dashed-dotted line and blue dashed line, 
respectively. Black solid line represents the surface brightness profile of 
the best-fit model. The bottom panel shows the residual. 
(b) Original image. 
(c) Best-fit model for the host galaxy without the nuclear component. 
(d) Residual image. All the images are shown with an asinh stretch.
\label{fig:fig2}}
\end{figure*}
\begin{figure*}[htp]
\ContinuedFloat 
\centering
\vspace{-5.0cm}
\includegraphics[width=180mm,page=2]{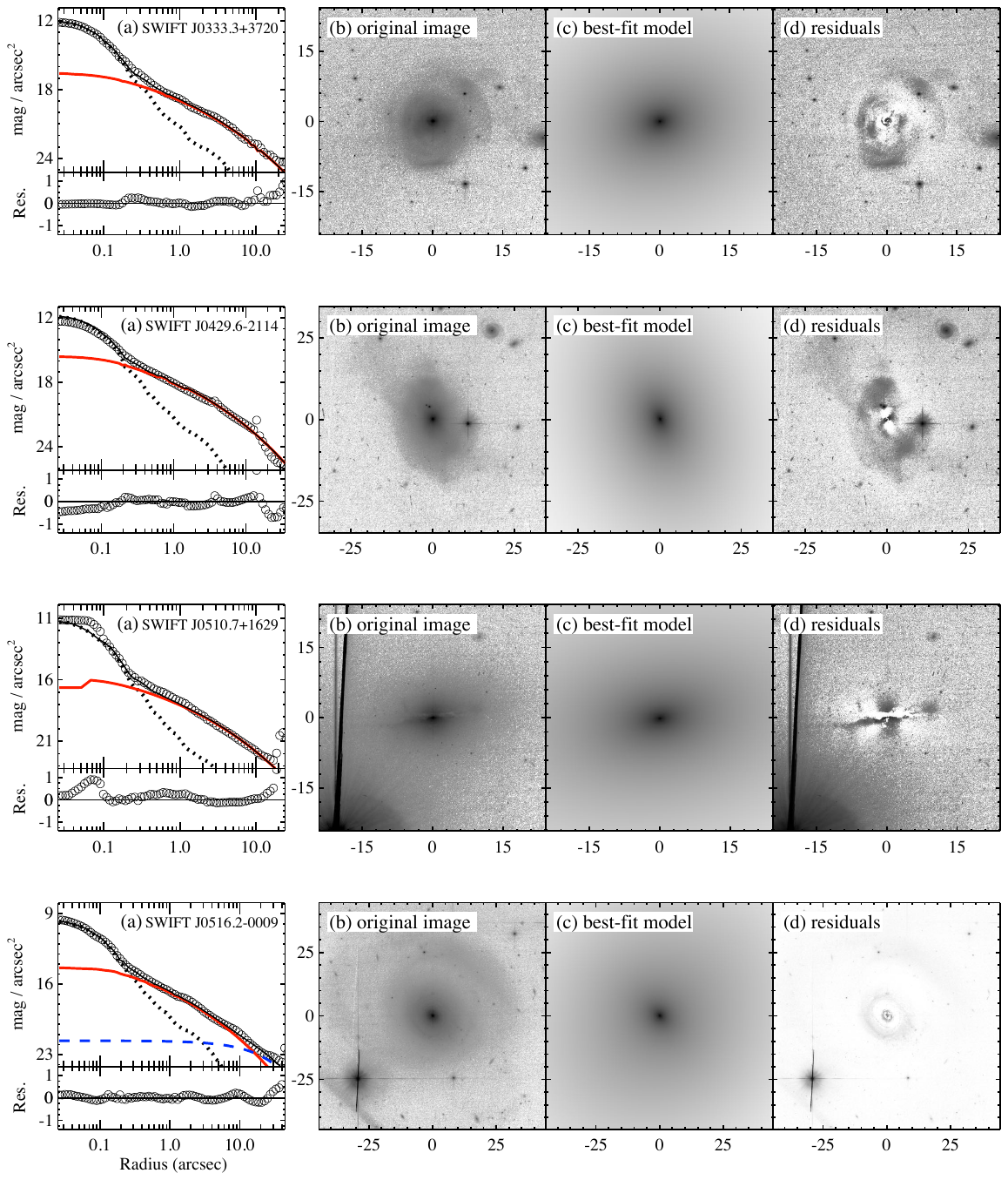}
\vspace{-1.5cm}
\caption{Continued}
\end{figure*}
\begin{figure*}[htp]
\ContinuedFloat 
\centering
\vspace{-5.0cm}
\includegraphics[width=180mm,page=3]{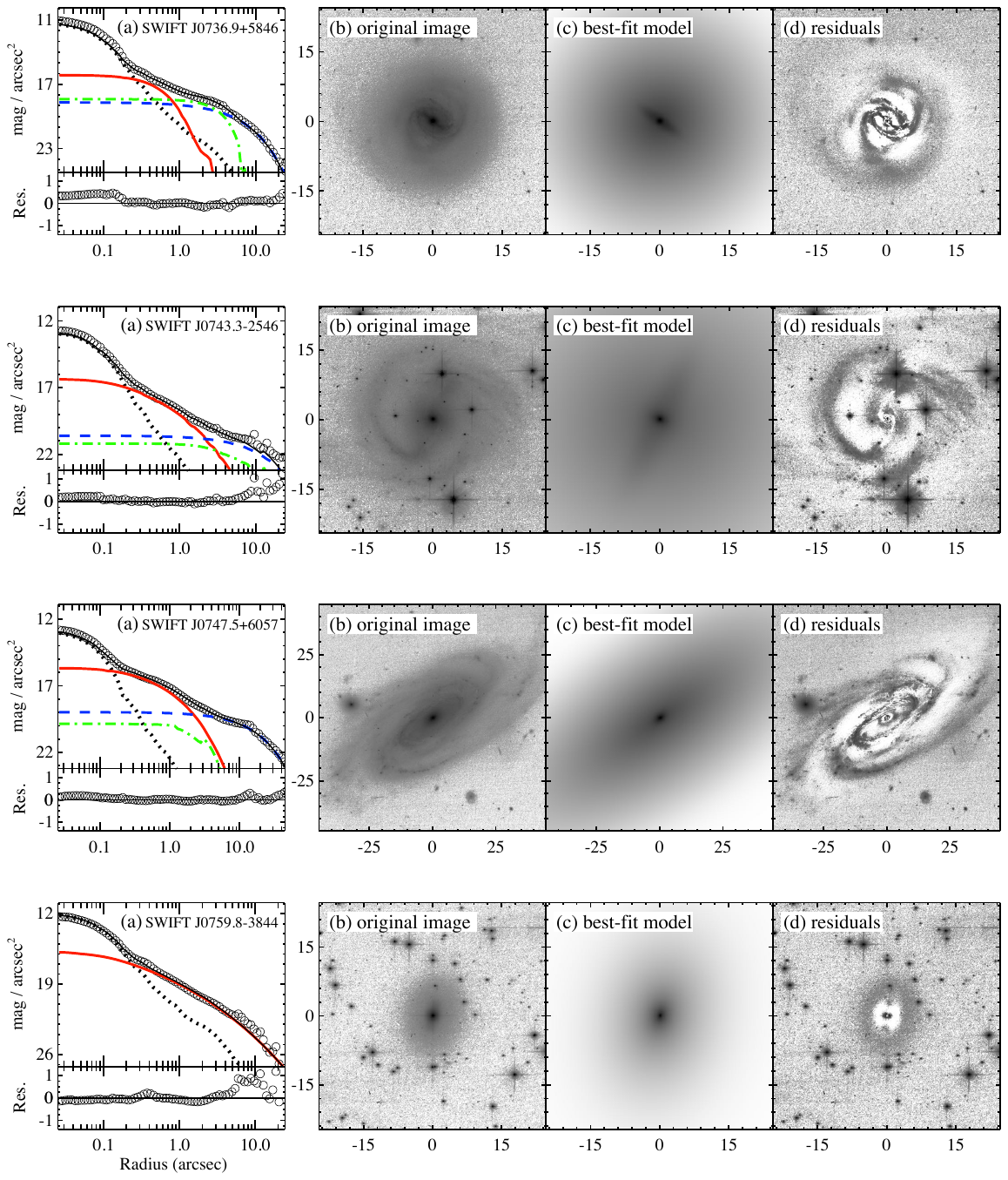}
\vspace{-1.5cm}
\caption{Continued}
\end{figure*}
\begin{figure*}[htp]
\ContinuedFloat 
\centering
\vspace{-5.0cm}
\includegraphics[width=180mm,page=4]{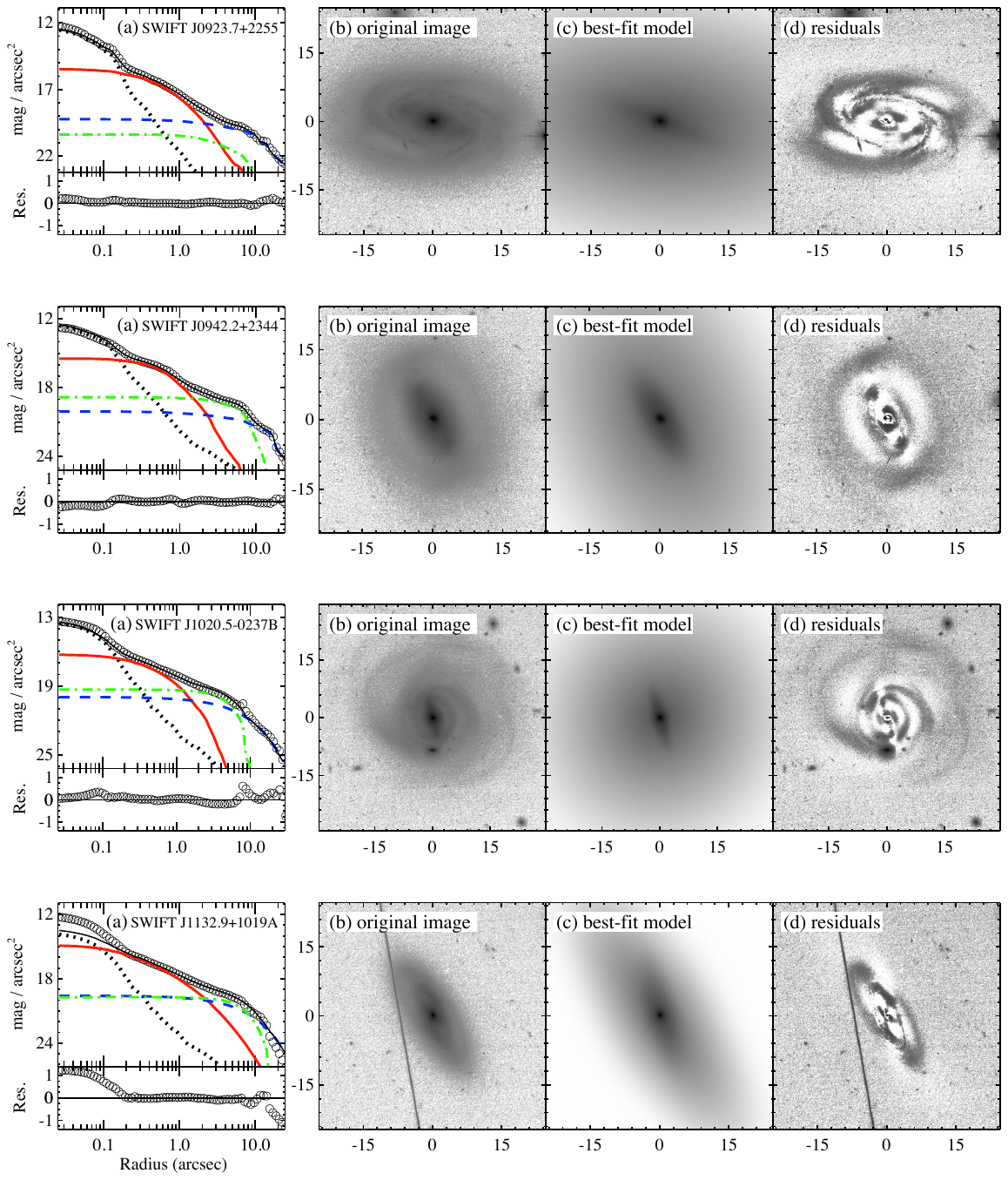}
\vspace{-1.5cm}
\caption{Continued}
\end{figure*}
\begin{figure*}[htp]
\ContinuedFloat 
\centering
\vspace{-5.0cm}
\includegraphics[width=180mm,page=5]{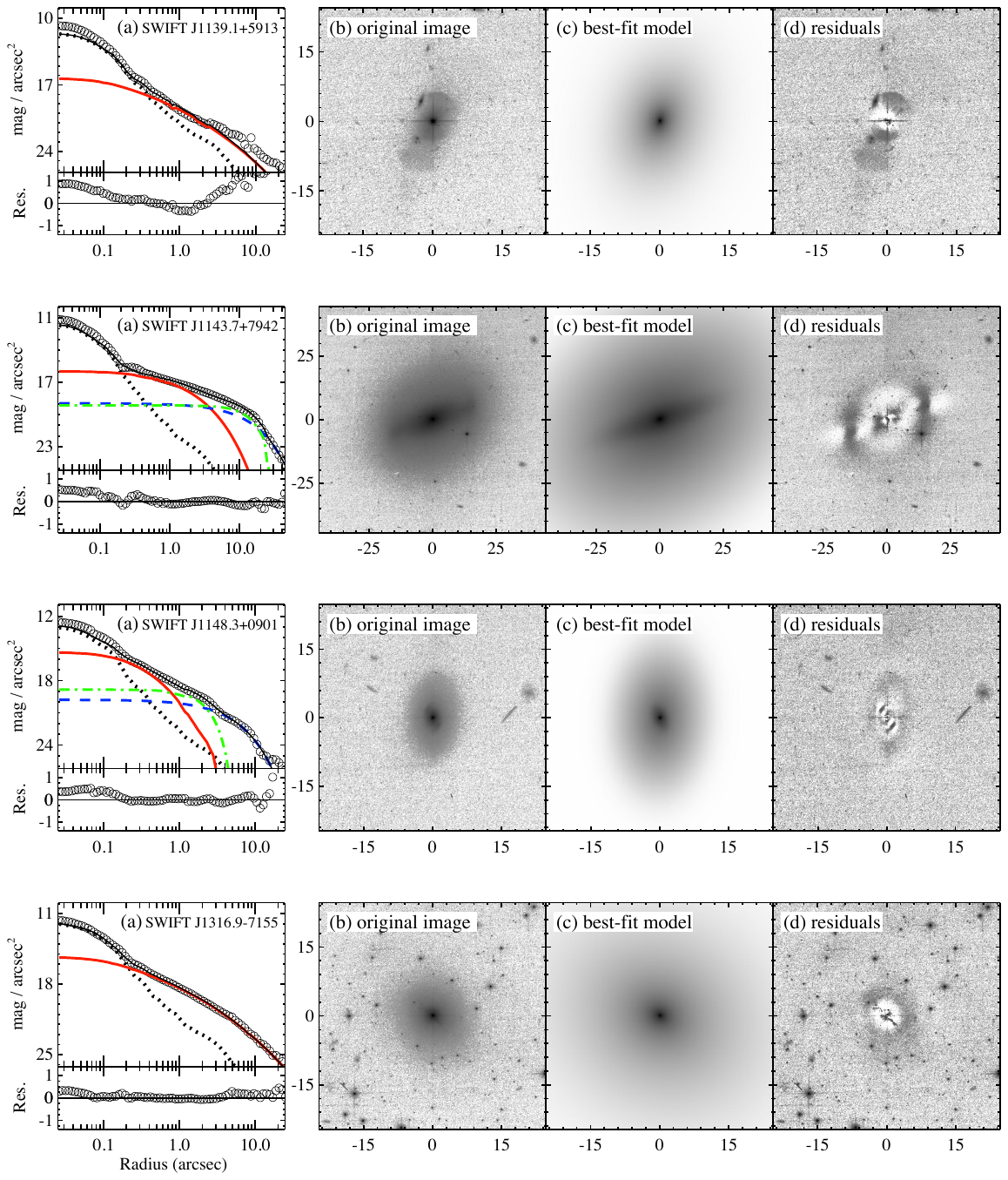}
\vspace{-1.5cm}
\caption{Continued}
\end{figure*}
\begin{figure*}[htp]
\ContinuedFloat 
\centering
\vspace{-5.0cm}
\includegraphics[width=180mm,page=6]{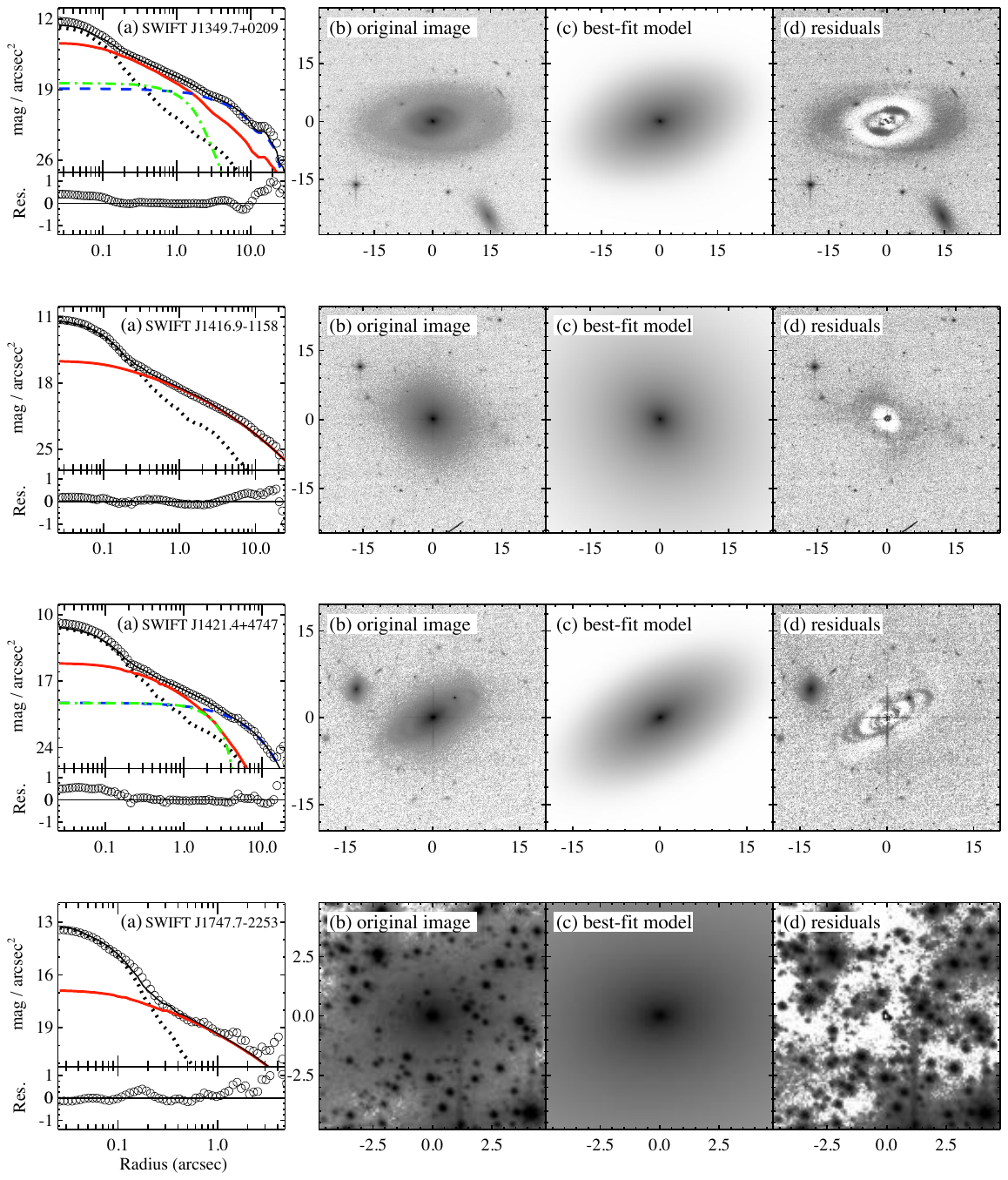}
\vspace{-1.5cm}
\caption{Continued}
\end{figure*}
\begin{figure*}[htp]
\ContinuedFloat 
\centering
\vspace{-5.0cm}
\includegraphics[width=180mm,page=7]{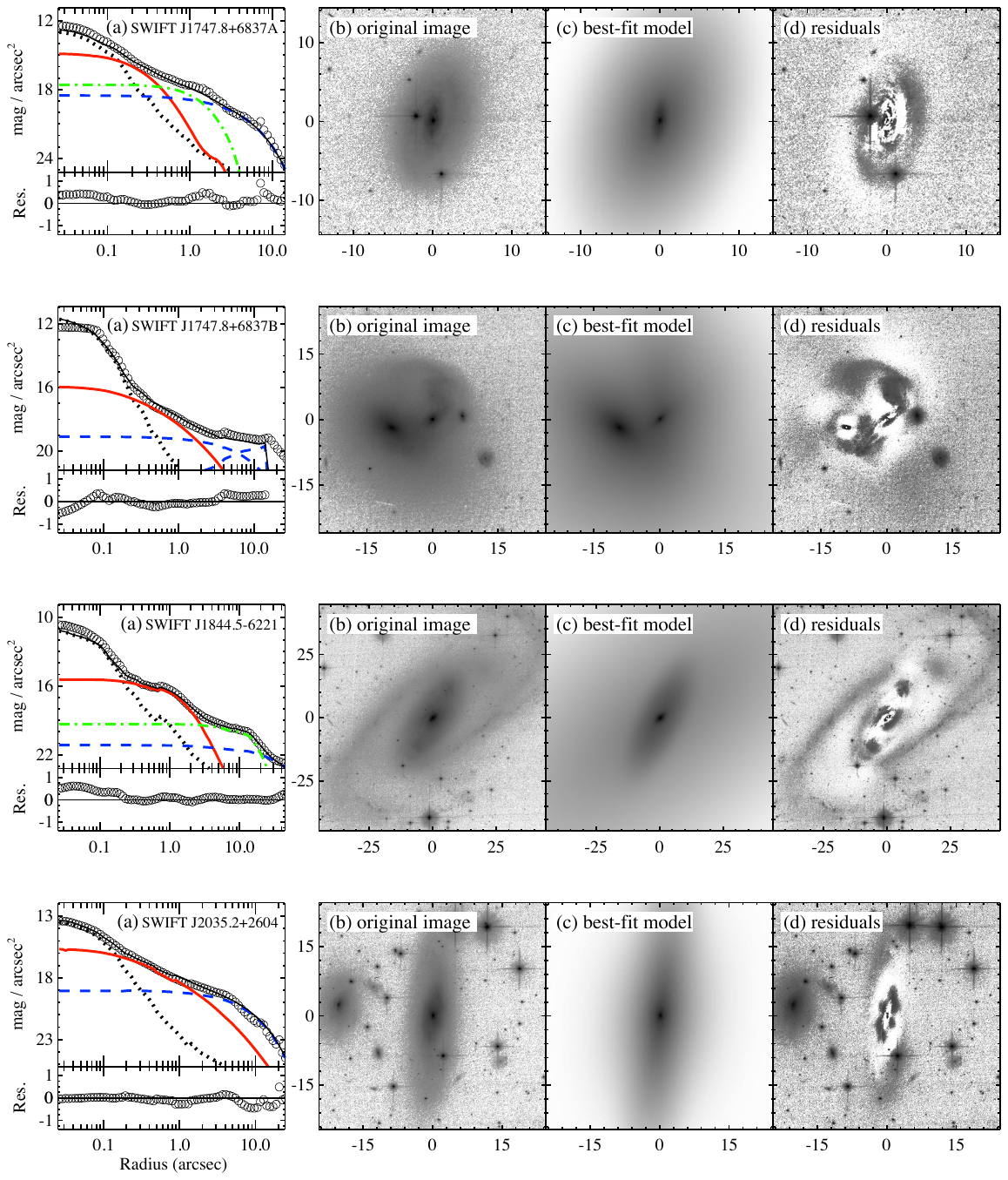}
\vspace{-1.5cm}
\caption{Continued}
\end{figure*}
\begin{figure*}[htp]
\ContinuedFloat 
\centering
\vspace{-5.0cm}
\includegraphics[width=180mm,page=8]{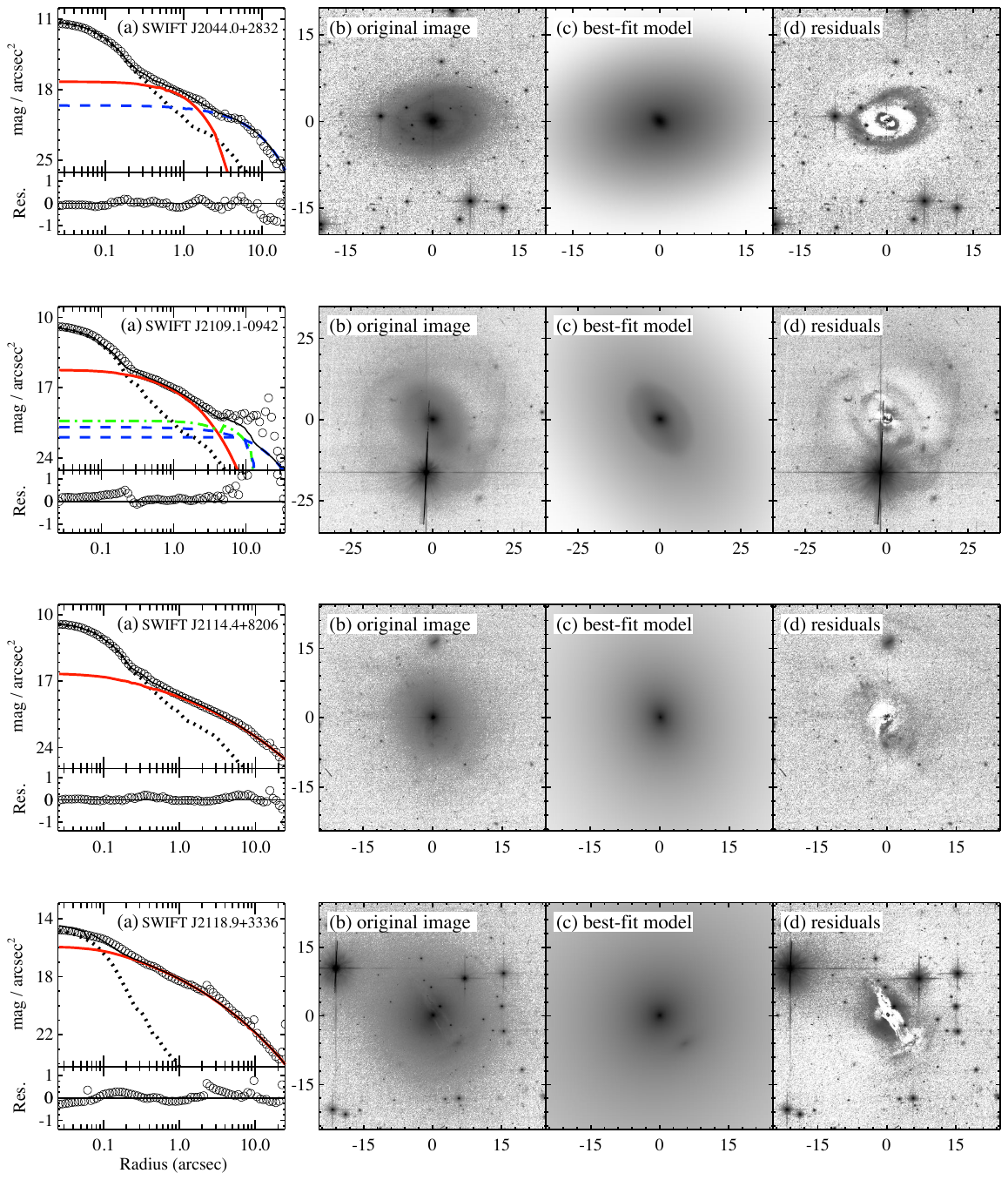}
\vspace{-1.5cm}
\caption{Continued}
\end{figure*}
\begin{figure*}[htp]
\ContinuedFloat 
\centering
\vspace{-5.0cm}
\includegraphics[width=180mm,page=9]{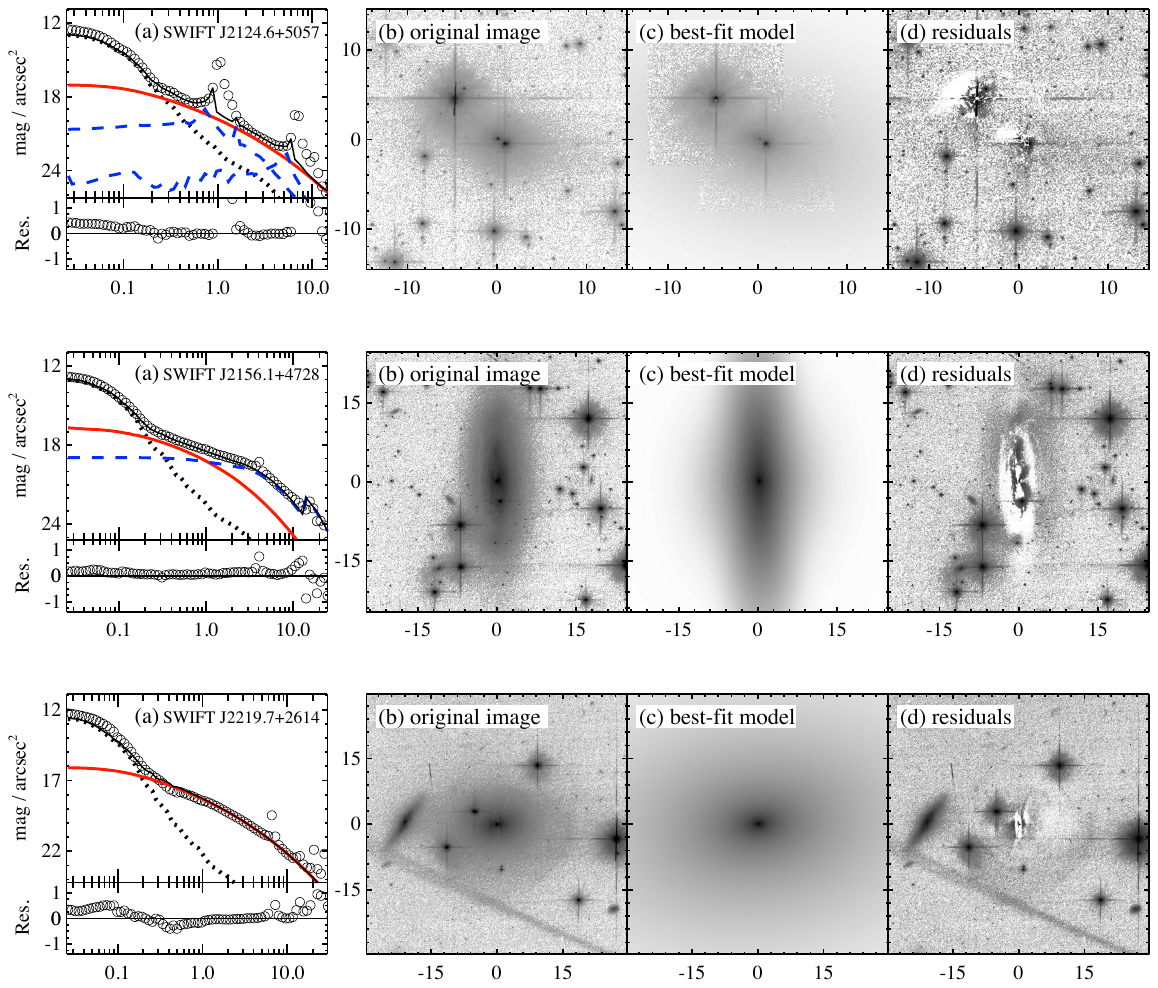}
\vspace{-7cm}
\caption{Continued}
\end{figure*}

\subsection{Image Decomposition}
To investigate the photometric properties of bulges in our sample, we 
performed a 2D imaging decomposition of 
the \hst\ images using \texttt{GALFIT} (\citealt{peng_2002, peng_2010}). 
Before conducting the fit, we manually masked neighboring objects 
and charge-bleeding trails due to saturation.  
For robust measurements, it is necessary to
adequately model non-bulge components, such as the nucleus, disk, 
bar, and oval if present along with the bulge component. We employed
the empirical PSF to approximate the nucleus. The sub-components
of the host galaxies were modeled with \ser\ profiles. 
As an initial test, we fitted the host galaxy with a single 
\ser\ profile, where we either fixed the \ser\ index ($n$) as 1 
or 4, or let it float as a free parameter.  
If there is no clear evidence of additional components 
(e.g., disk or bar) in the residual images, we conclude that the host 
is well fitted with the single \ser\ component (i.e., elliptical). 
For a sanity check for the goodness of the fit, we also 
utilized 1D surface brightness profiles of the original images and 
best models and their residuals (Fig. \ref{fig:fig2}).

If any structures other than the bulge are revealed in the residual images, 
we refit the imaging data by adding \ser\ components: a
\ser\ component with $n=1$ for an exponential disk, a
\ser\ component with $n=0.5$ for a bar, and a \ser\ component with free $n$ 
for an oval. 
If the fitting result is severely affected by the light from companion 
objects, we simultaneously fit them with \ser\ profiles for extended sources
or PSFs for point sources. 
Finally, Fourier modes were additionally implemented to 
account for the lopsidedness of the host if the host galaxy is heavily 
disturbed. The detailed fitting results, including 1D surface brightness 
profiles, are displayed in Figure \ref{fig:fig2}.

Owing to the bright nucleus and PSF mismatch, it is occasionally difficult
to choose the best model for the host galaxy from the decomposition with 
\texttt{GALFIT}. In that case, the fitting result with the most physically meaningful
parameters and a smaller reduced $\chi^2$ value was selected as the 
best fit. For example, \ser\ models with a large \ser\ index ($n>8$) or a 
tiny effective radius ($R_{\rm e} \leq$ a few pixels) were not considered as 
reliable. 

To estimate the uncertainty of the bulge luminosity, we adopted the recipe 
proposed by \citet{kim_2017} based on imaging simulations. The 
uncertainty of the bulge luminosity was initially determined by the 
bulge-to-nucleus light ratio 
(0.3 mag for $L_{\rm bul}$/$L_{\rm nuc} \geq 0.2$ and 0.4 mag for 
$L_{\rm bul}$/$L_{\rm nuc} < 0.2$). Then, an additional component (e.g., disk)
for the host would increase the uncertainty by 0.1 mag. 
 Additional uncertainty can be introduced from imperfect sky subtraction,
which was done as part of the drizzling process. Because the field of view is sufficiently large for robust sky determination, the measurement error for the sky estimation is expected to be small. Nevertheless, to quantify the effect of sky uncertainty, we perform the decomposition by fixing the sky to $1\sigma$ and $-1\sigma$ of the original sky value. This experiment often resulted in host components with unreliable structural parameters (e.g., extremely small or large \ser\ indices with large effective radii), but the impact on the bulge luminosity is modest, usually only within 0.2 mag. 
Note that PSF mismatch appears to be severe in a few targets (e.g., SWIFT J0123.9$-$5846, SWIFT J1132.9+1019A). However, because the bulges are large and bright, as shown in the radial profiles, the total bulge luminosity is only affected at the level of $\sim0.2-0.3$ mag. 
In addition, such PSF mismatch was taken into account when estimating the error budget (\citealt{kim_2017}).

We applied a correction for Galactic extinction (\citealt{schlafly_2011}). 
We performed $k-$correction using the method in \citet{chilingarian_2010}
and \citet{chilingarian_2012}. The color information of host galaxies 
is unknown. As a result, we adopt $V-I_{\rm C}$ color inferred from the 
morphological types of hosts (\citealt{fukugita_1995}), which are 
determined from visual inspection. 
For the nucleus, $V-I_{\rm C}\sim0.6$ is calculated from the QSO 
composite spectrum (\citealt{vandenberk_2001}). The conversion of 
$M_{\rm F814W}$ to 
$M_I$ was not applied because the conversion factor is known to be 
negligible ($<0.05$ mag; \citealt{fukugita_1995}, \citealt{harris_2018}).
Table \ref{tab:table3} lists the photometric properties of the sample obtained through 
the imaging decomposition.

\section{Results}

\subsection{\mlr\ Relation}
To estimate the \mlr\ relation for our sample, we employed 
a $\chi^2$ minimization 
fit (modified FITEXY) given in \citet{tremaine_2002}, which accounts for 
the measurement errors in both $M_{\rm BH}$ and $M_{I,\rm bul}$ and intrinsic 
scatter ($\epsilon_0$).
We fit the \mlr\ relation of the form 
\begin{eqnarray}
\log (M_{\rm BH}/M_{\odot})=\alpha\,M_{I,\rm bul}+\beta.
\end{eqnarray}
Given the small sample size, we fixed the slope to $\alpha=-0.57$, which was 
derived from the inactive galaxies, and only solved for the zero point 
($\beta$). 
To compute the \mlr\ relation of the inactive galaxies in $I$-band from that 
in $K$-band (\citealt{kormendy_2013}), we used mass-to-light-ratio in 
$K$- and $I$-band inferred from $B-V$ color (\citealt{into_2013}).

\begin{figure*}[h]
\centering
\includegraphics[width=75mm]{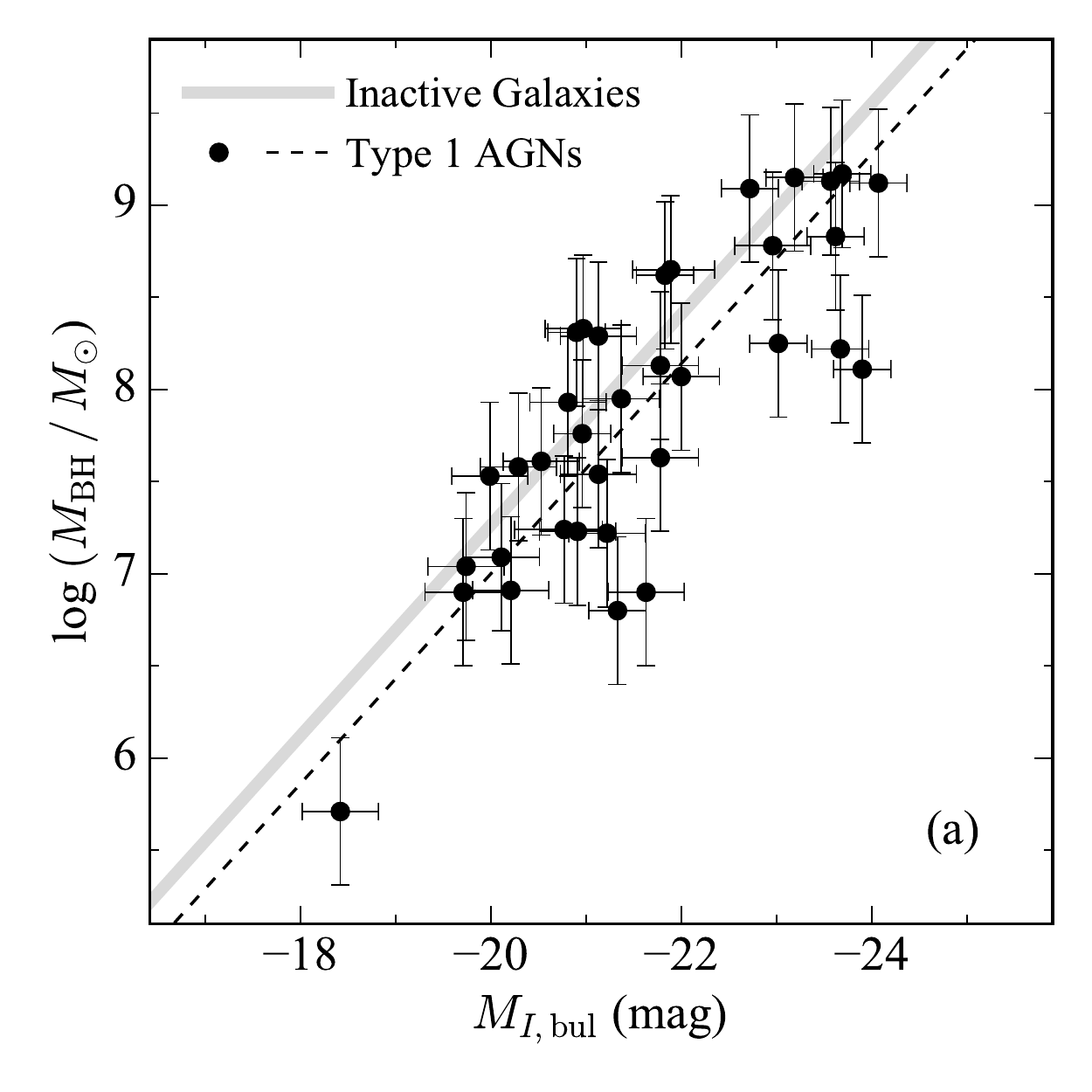}
\includegraphics[width=75mm]{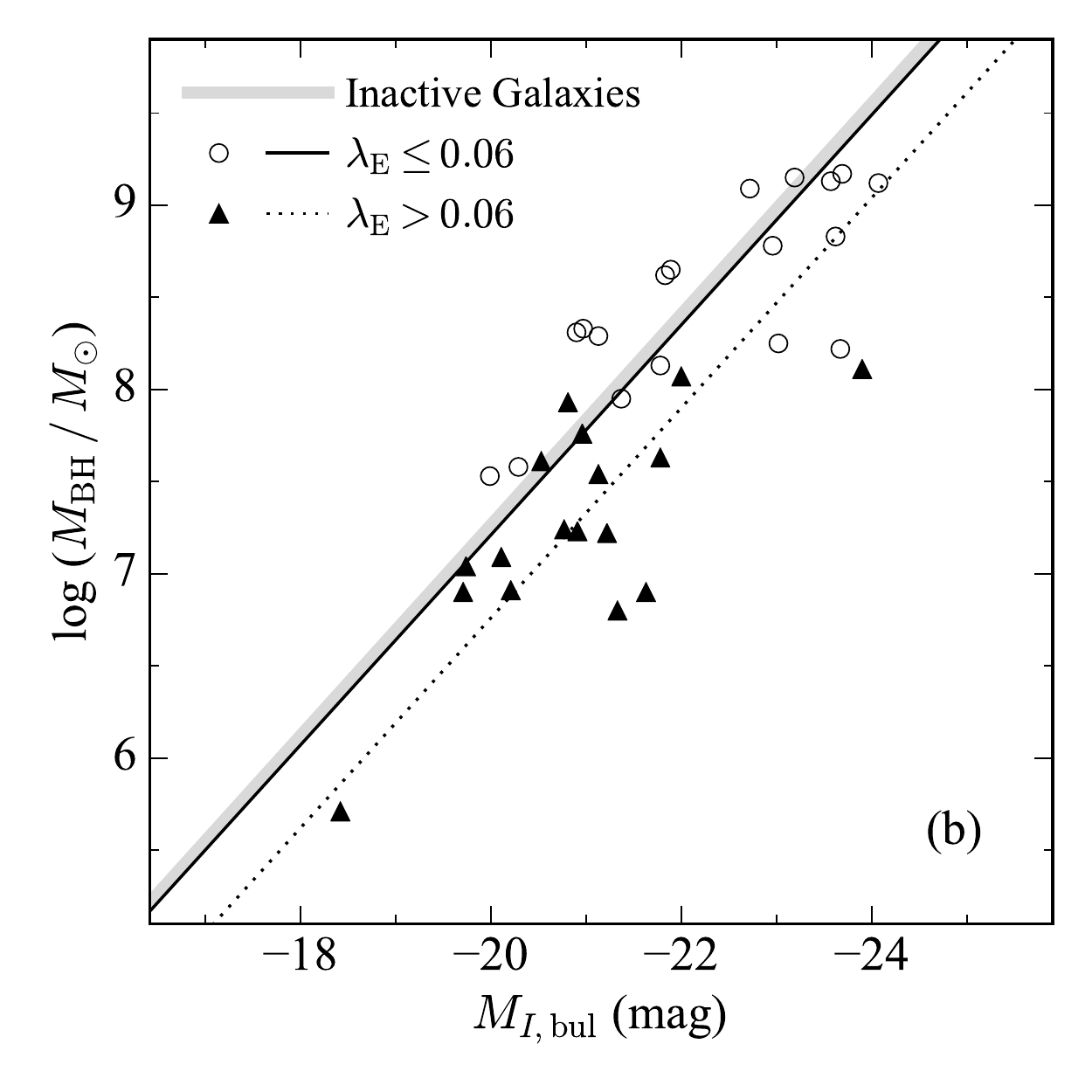}
\includegraphics[width=75mm]{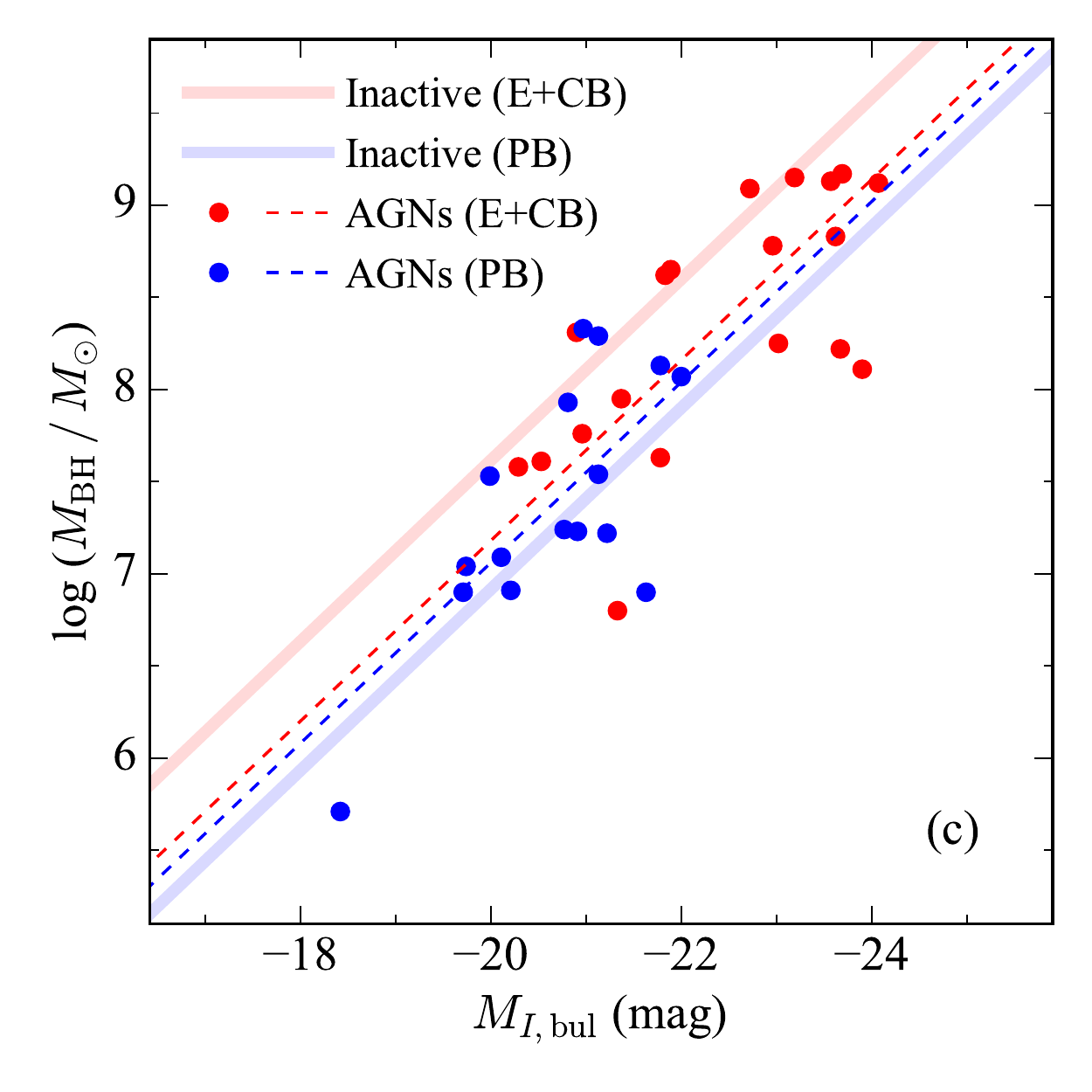}
\includegraphics[width=75mm]{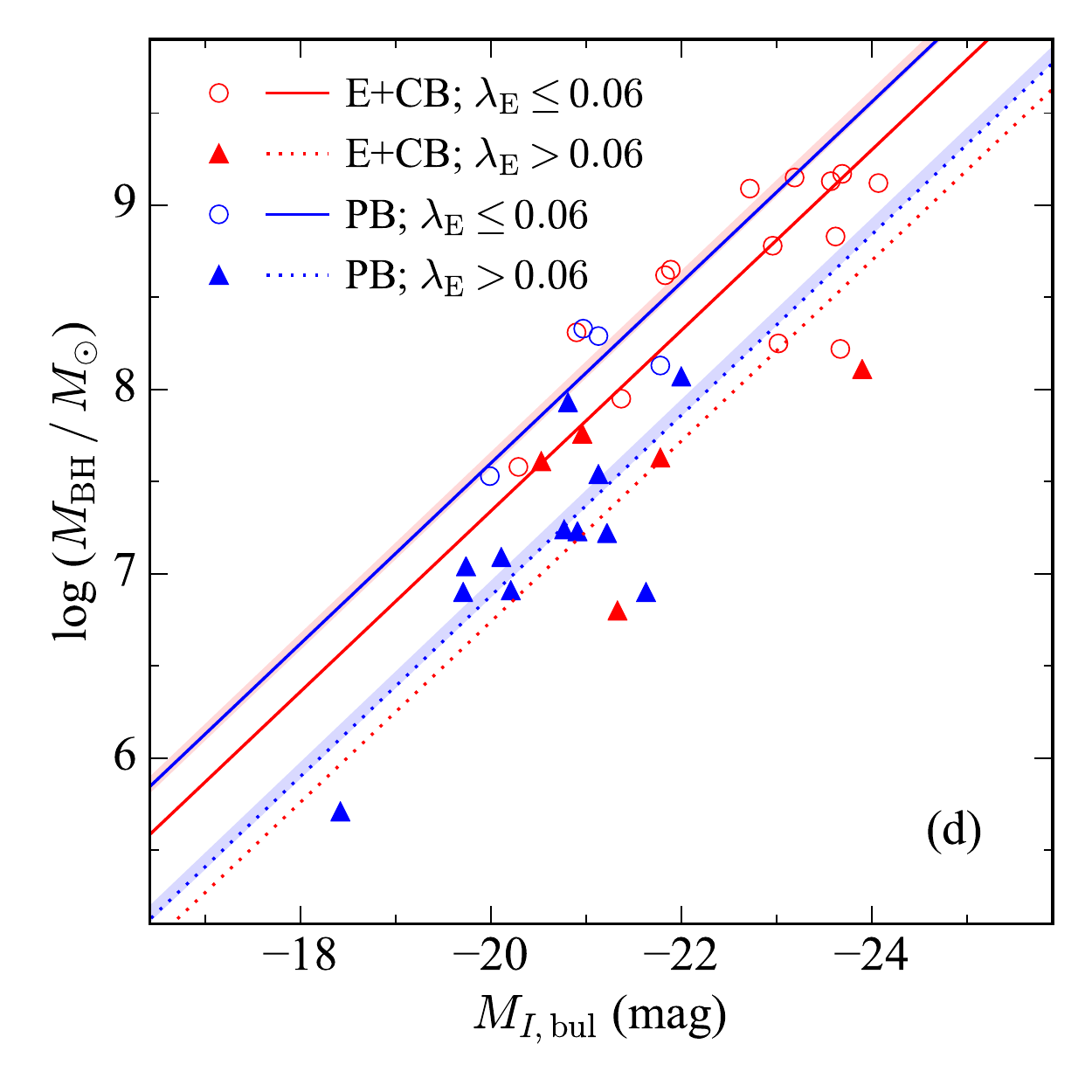}
\caption{Relations between BH mass ($M_{\rm BH}$) and absolute $I$-band bulge 
magnitude ($M_{I, \rm bul}$) for type 1 AGNs. In all panels, shaded line 
denotes the relation of inactive galaxies from \citet{kormendy_2013}.  
(a) Filled circles and the dashed line denote type 1 AGNs and their relation. 
(b) Open circles (filled triangles) and solid line (dashed line) represent 
AGNs with low (high) Eddington ratio and their relation, respectively. 
(c) The sample is divided into two subgroups according to bulge types. 
Blue shaded (red shaded) line denotes the relation of pseudo-bulges 
(ellipticals and classical bulges) in inactive galaxies. Red (blue) circles 
and dashed line denotes AGNs with ellipticals and classical bulges
(pseudo-bulges) and their relation, respectively. 
(d) Red (blue) circles and solid line denote ellipticals and classical 
bulges (pseudo-bulges) with low Eddington ratio in our sample and their 
relation, respectively. Red (blue) triangles and dotted line denote 
ellipticals and classical bulges (pseudo-bulges) with high Eddington ratio in 
our sample and their relation, respectively.
\label{fig:fig3}}
\end{figure*}

It appears that the AGNs in our sample tend to have a lower zero point 
compared to inactive galaxies (Fig. \ref{fig:fig3}). To find the main driver of this 
offset, we divided the sample into two subgroups: one with a lower 
Eddington ratio than the median value ($0.06$) and one with a higher 
Eddington ratio than the median value. We find that AGNs with 
larger Eddington ratio systematically lie below those with smaller Eddington ratio in the \mlr\ relation, indicating that the 
zero point depends on the Eddington ratio. 

However, given the fact that hosts with classical bulges and pseudo-bulges 
follow a different \mlr\ relation (\citealt{kormendy_2013}), it is 
essential to 
assess this trend by dividing the sample according to the bulge type 
(\citealt{ho_2014}). 
In this light, we performed the bulge classification based on the \ser\ index 
($n$) and $B/T$. Following the methods given in \citet{fisher_2008}, 
\citet{gadotti_2009} and \citet{gao_2020}, bulges with $n>2$ and $B/T>0.2$ 
($n\leq2$ or $B/T\leq0.2$) are classified as a classical bulge 
(pseudo-bulge). Furthermore, we computed the \mlr\ relation of inactive 
galaxies for different bulge types utilizing the data from 
\citet{kormendy_2013}. The \mlr\ relation of the pseudo-bulges has a 
substantially larger
scatter ($\epsilon_0\sim0.63$ dex for the pseudo-bulge in inactive galaxies) 
compared to the classical bulges ($\epsilon_0\sim0.31$ dex), which can 
introduce a systematic bias in the comparison between the two bulge 
types (Tab. \ref{tab:table4}). Therefore, classical bulges and ellipticals are considered only for 
further analysis. By fixing the slope to that of classical 
bulges and ellipticals in inactive galaxies ($\sim-0.49$), we again find 
that AGNs with classical bulges or ellipticals lie systematically below the 
relation of inactive galaxies and the zero point offset is lower for 
objects with greater Eddington ratio. This observation confirms our previous 
finding for the entire sample, that the zero point depends on 
the Eddington ratio.

\subsection{Kormendy Relation}
There is an anti-correlation between the effective radius ($R_{\rm e}$) and 
the mean surface brightness (\mue) within the effective radius, known as 
the Kormendy relation (\citealt{kormendy_1977}). Here, we present the 
Kormendy relation for our sample. \citet{fisher_2008}, 
\citet{gadotti_2009}, and \citet{gao_2020} reported that classical bulges 
and pseudo-bulges are distinctive in the Kormendy relation, while 
classical bulges follow the Kormendy relation similar to that of 
ellipticals. Therefore, we investigated the Kormendy relations as
a function of bulge type. We fit the Kormendy relation of the form 
\begin{eqnarray}
\langle \mu_{\rm e} \rangle = \kappa \log (R_{\rm e}/{\rm kpc})+\gamma,
\end{eqnarray}
where \mue\ is the $I$-band mean surface brightness in the units of mag arcsec$^{-2}$. 
We adopted the ordinary least squares bisector method for the fit as both 
parameters are independent (\citealt{isobe_1990}). 

\begin{figure*}[h]
\centering
\includegraphics[width=75mm]{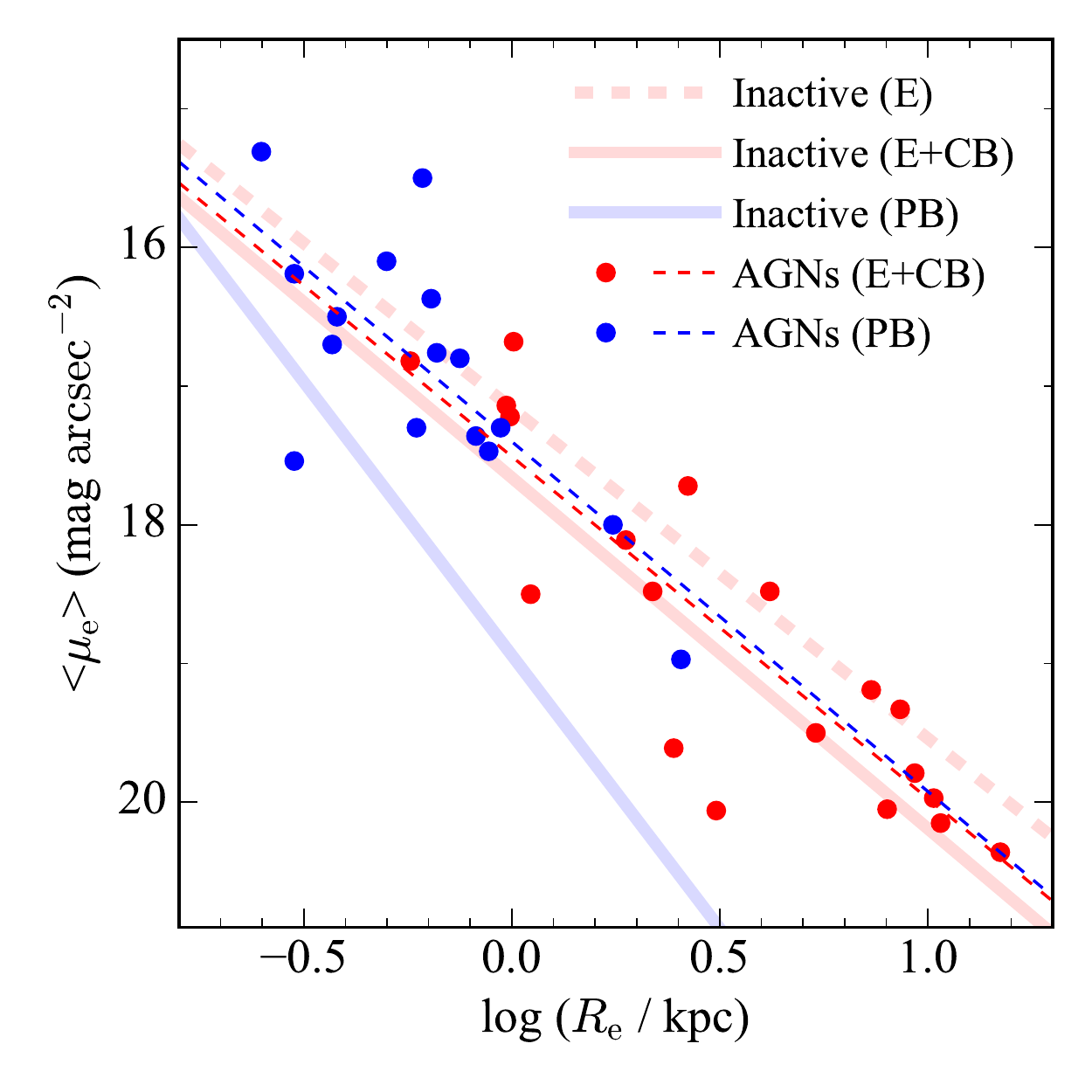}
\includegraphics[width=75mm]{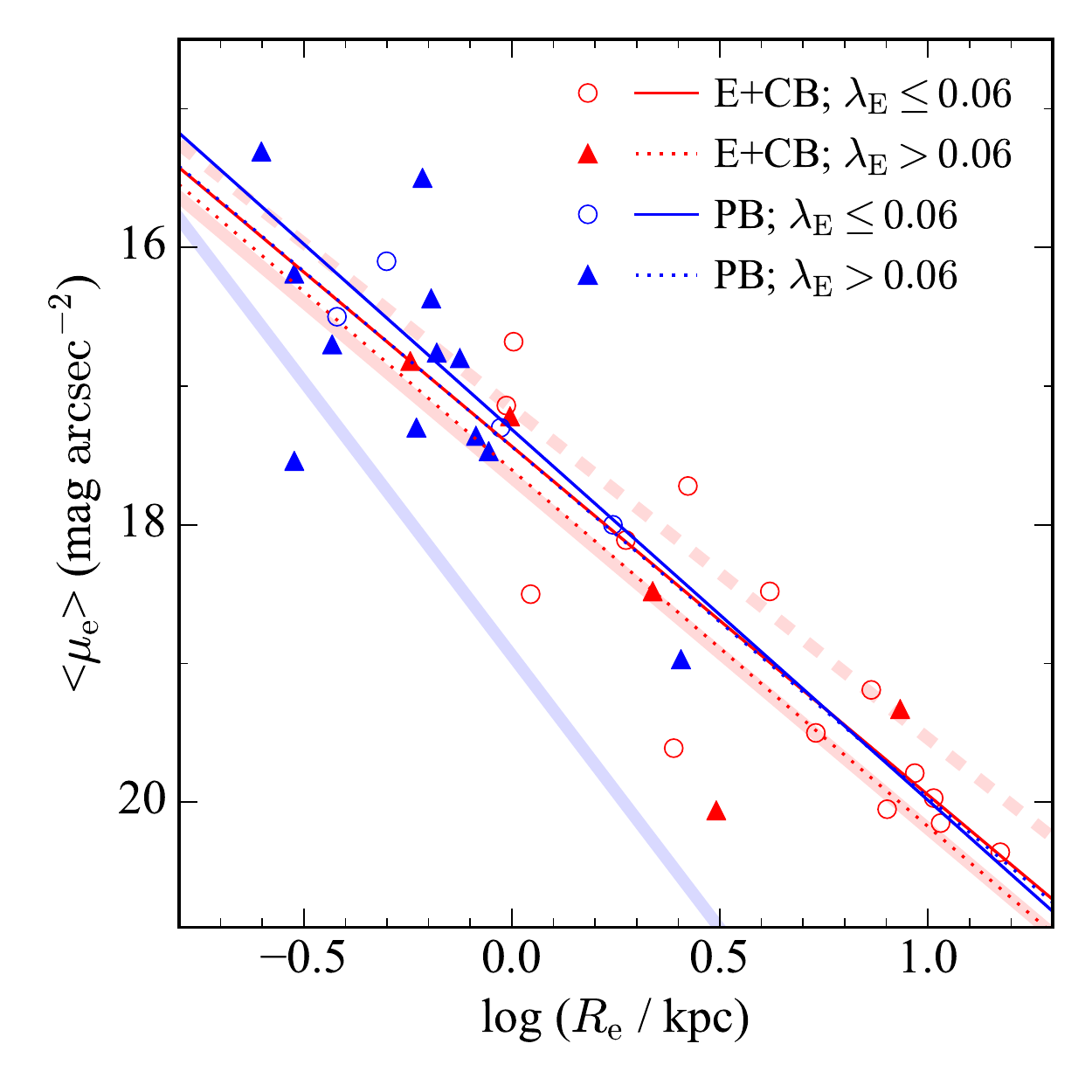}
\caption{Relation between effective radius ($R_{\rm e}$) and mean surface 
brightness ($\langle\mu_{\rm e}\rangle$) within $R_{\rm e}$ for type 1 AGNs. 
The sample is divided into two subsamples according to bulge types (left) 
and four subgroups according to bulge types and Eddington ratio (right). 
For both panels, red (blue) symbols and lines denote ellipticals and 
classical bulges (pseudo-bulges). Kormendy relations for inactive galaxies 
adapted from \citet{kim_2019} are shown by thick shaded lines. 
Kormendy relation for elliptical galaxies from \citet{gao_2020} is
denoted by thick red dashed line. In the left 
panel, red and blue dashed lines represent the Kormendy relations for AGNs 
with ellipicals and classical bulges and those with pseudo-bulges,
respectively. In the right panel, AGNs with low (high) Eddington ratio 
and their Kormendy relation are denoted by open circles (filled triangles) 
and solid (dotted) lines, respectively.
\label{fig:fig4}}
\end{figure*}

The Kormendy relation of inactive galaxies in the $I$ band is inferred from 
that in $R$ band given in \citet{kim_2019}. This process is done by 
assuming $R-I=0.65$ for ellipticals and classical bulges, which is 
equivalent to the value for a galaxy of Hubble type Sab, and 
$R-I=0.57$ for pseudo-bulges, which is equivalent to the value for an
Scd spiral (\citealt{fukugita_1995}). For the comparison, we also 
adopted the Kormendy relation of inactive ellipticals in the $I$ band. 
It was converted from that in $R$ band from \citet{gao_2020} by 
assuming $R-I=0.70$ of an elliptical galaxy (\citealt{fukugita_1995}).
Figure \ref{fig:fig4} shows that, for ellipticals and classical bulges, there is 
little difference between the AGNs in our sample and those in inactive 
galaxies. However, for pseudo-bulges, the AGNs in our sample appear to 
systematically have brighter \mue\ compared to those in inactive galaxies at 
a given \re. In both types, we found no evidence for a dependence of 
the Kormendy relation on Eddington ratio. The detailed fitting results 
for different subsamples are summarized in Table \ref{tab:table5}.

\section{Discussion}

\subsection{Origin of the Offset in the \mlr\ Relation}
 We find that AGNs in our sample have a lower zero point compared to 
the inactive galaxies. Additionally, the magnitude of the offset in the 
zero point ($0.4-0.8$ dex) is anti-correlated with the Eddington ratio. 
This finding can be interpreted in three ways. 
For AGNs with a high Eddington ratio, either the bulges are overluminous 
compared to those in inactive galaxies, the BHs are less massive than those 
in inactive galaxies at a given bulge luminosity, or BH masses are 
somehow underestimated. In contrast, ellipticals and classical bulges in 
our sample follow the Kormendy relation similar to that in inactive galaxies, 
thereby revealing that there is no excess in the bulge luminosity. Therefore, 
the offset in the \mlr\ relation is unlikely to occur due to the 
overluminous bulge. 

As a BH is actively growing during the AGN phase, it can be naturally expected 
that AGNs can systematically have less massive BHs than inactive galaxies
at a given bulge luminosity. Assuming a constant Eddington ratio
within the AGN lifetime ($t_{\rm AGN}$), a BH growth factor ($\frac{M_{\rm final}}{M_{\rm int}}$) 
during the AGN phase can be expressed as 
\begin{eqnarray}
\frac{M_{\rm final}}{M_{\rm int}} \approx 
\exp(\lambda \frac{1-\eta}{\eta} \frac{t_{\rm AGN}}{t_{\rm Edd}}),
\end{eqnarray}
where $M_{\rm final}$ is the final BH mass after the AGN phase, 
$M_{\rm int}$ is the initial BH mass,
$\eta$ is the radiative efficiency of the accretion disk and 
$t_{\rm Edd}$ is the Eddington time scale ($\approx 0.45$ Gyr; 
\citealt{volonteri_2005}). For simplicity, we assume $\eta\sim0.1$ and 
$t_{\rm AGN}\sim 0.05$ Gyr (\citealt{yu_2002, martini_2004, kim_2019}). 
By adopting the median Eddington ratio of our sample 
($\lambda_{\rm E}\sim0.06$), the BH growth factor during the AGN phase 
($\sim1.06$ or 0.03 dex) is almost negligible compared to the offset 
($\sim 0.4$ dex) in the \mlr\ relation. Even with an extreme 
assumption ($\eta\sim0.01$; e.g., \citealt{davis_2011}), the BH 
growth factor is less than 0.3 dex. This reveals that undermassive BHs 
are unlikely to be the main driver of the offset. 

\begin{figure*}[h]
\centering
\includegraphics[width=75mm]{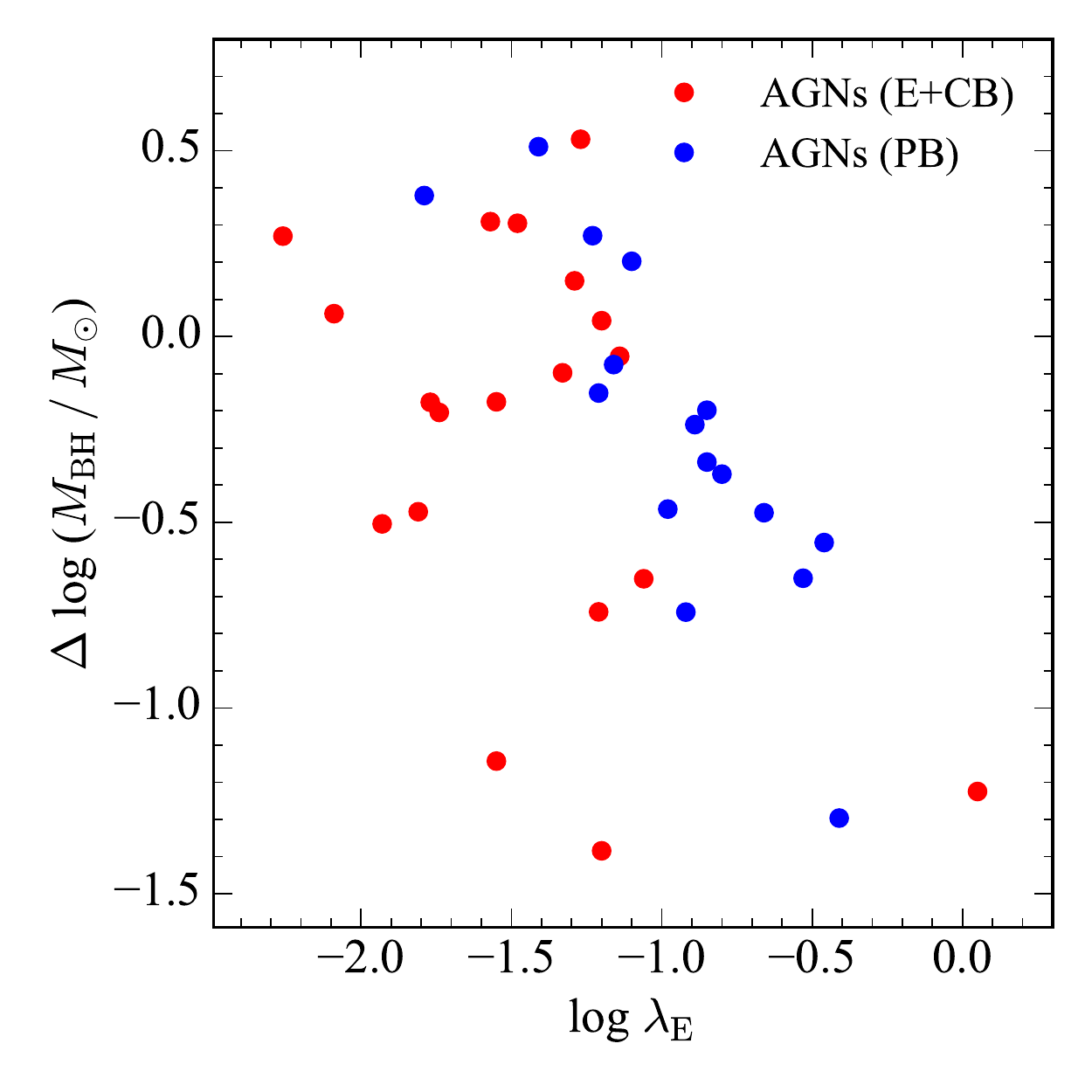}
\caption{Comparison of Eddington ratio ($\lambda_{\rm E}$) with BH mass 
offset ($\Delta \log M_{\rm BH}$) from the \mlr\ relation of inactive 
galaxies. BH mass offset of type 1 AGNs is calculated by subtracting the 
virial BH mass estimates from that inferred from the \mlr\ relation of 
inactive galaxies at a given bulge luminosity. Type 1 AGNs hosted by
ellipticals and classical bulges are plotted as filled red circles, while 
those hosted by pseudo-bulges are plotted as filled blue circles.
\label{fig:fig5}}
\end{figure*}

It has long been suggested that the scaling factor in the virial mass
estimator for BHs can be sensitive to physical properties of AGNs 
(e.g., Eddington ratio, bolometric luminosity and inclination of BLR; 
\citealt{marconi_2008, ho_2014, mejia_2017}). Therefore, the dependence 
of the zero point offset on the Eddington ratio can be naturally attributed 
to the different scaling factors instead of a single universal value. To 
assess this effect more quantitatively, we calculated $\Delta M_{\rm BH}$ in 
each object, which is defined as the offset between the BH 
mass estimated from the virial method and that inferred from the 
inactive \mlr\ relation at a given bulge luminosity in a log-log space. 
Then we compared $\Delta M_{\rm BH}$ with the Eddington ratio (Fig. \ref{fig:fig5}). 
The Spearman correlation coefficient is $-0.49$ ($p$-value $\sim 0.003$), 
which implies that the correlation between two variables is significant.
To validate the statistical significance of the correlation, we employed 
the bootstrapping resampling method. From this experiment, we found a 68\% 
confidence interval of $-0.45$ to $-0.64$, 
again indicating that the correlation is statistically significant.
Additionally, $\Delta M_{\rm BH}$ appears to converge to 0 when the 
Eddington ratio is smaller than 0.05, which supports the idea that 
the zero point offset is due to the dependence of the scaling factor 
on the Eddington ratio.
One may argue that the secondary parameter (e.g., $M_{\rm BH}$) can be 
another origin of the systematic offset. To test this hypothesis, we computed 
the Spearman correlation coefficient ($\sim0.41$ with $p$-value $\sim 0.02$) between $\Delta M_{\rm BH}$ and $M_{\rm BH}$. Although the coefficient value is slightly 
smaller than that of the correlation between $\Delta M_{\rm BH}$ and the Eddington ratio, it appears to be moderate. We perform a partial correlation analysis to further investigate 
whether Eddington ratio or $M_{\rm BH}$ is the primary parameter driving the zero point offset. We find a higher partial correlation coefficient ($\rho \sim -0.38$) between $\Delta M_{\rm BH}$ and Eddington ratio after removing the contribution from BH mass. Note that the coefficient between $\Delta M_{\rm BH}$ and BH mass excluding the effect from Eddington ratio is $-0.13$. Therefore, 
we conclude that Eddington ratio is likely to be a main driver of the zero point 
offset. The last caveat is that low-mass AGNs ($M_{\rm BH}\leq 10^{7.5}M_{\odot}$) with a low Eddington ratio may
be excluded with the flux-limited hard X-ray selection. This may be a possible 
scenario to explain the observed zero point offset, but we cannot address it in this study given the small sample size.

\subsection{Comparison with Previous Studies}
Previous studies with \hst\ imaging dataset of nearby type 1 AGNs 
also utilized the BH$-$host and Kormendy relations to investigate the
physical connection between SMBHs and host galaxies of AGNs 
(e.g., \citealt{kim_2008b,kim_2019,zhao_2019,zhao_2021}).
However, they reached somewhat different conclusions. For example, 
\citet{kim_2019} found that AGNs lie systematically below in the \mlrr\ 
relation compared to normal galaxies, and they concluded that bulges hosting AGNs 
are overluminous possibly due to the young stellar population based on the 
archival data of nearby AGNs ($z<0.35$). Those observational results are in 
broad agreement with our finding in the sense that AGNs are offset from the 
normal galaxies in the \mlb\ relation. However, their interpretation is 
somewhat different from ours, possibly because they partly used $B$- and 
$V$-band magnitudes that are more sensitive to young stellar populations 
compared to $I$-band used in this study. Additionally, unlike our sample, 
the sample in \citet{kim_2019} is somewhat heterogeneous, and the 
archival imaging data were obtained in various observing configurations 
(e.g., instrument, filter, and exposure time). Therefore, their result 
can suffer from hidden biases. 

In contrast, \citet{zhao_2021} reported that AGNs with massive hosts 
follow scaling relations (\mmr\ and Kormendy relation) similar to those
of normal galaxies, based on color images of PG quasars ($z<0.5$). 
These results are inconsistent with our finding that ellipticals and 
classical bulges hosting AGNs deviate from the \mlr\ relation of those in 
normal galaxies. The massive AGNs in \citet{zhao_2021} are more distant 
($z\sim0.4$) than those in our sample ($z<0.1$), and may be affected by 
cosmic evolution and/or unknown bias. They also found that only the 
late-type galaxies with pseudo-bulges of AGNs deviate in the scaling 
relations, indicating that the brightness of pseudo-bulges hosting AGNs is 
enhanced due to recent star formation (e.g., \citealt{zhuang_2020,
xie_2021, zhuang_2021}). 
Interestingly, we also find 
the same offset in the Kormendy relation only for pseudo-bulges.

\section{Conclusion}
In this paper, we estimated the photometric properties of the bulges in 
the host galaxies by performing the careful imaging decomposition on 
\hst\ images of 35 nearby AGNs, originally selected from hard X-ray 
data. Along with BH mass estimates from the virial methods, we examined 
the \mlr\ and Kormendy relations of our sample and compared them with those 
of inactive galaxies. The main results from these experiments can be
summarized as follows.

\begin{itemize}
    \item The \mlr\ relation of our sample AGNs slightly deviate from that 
    of inactive galaxies in the sense that the BH mass in our sample AGNs 
    is $\sim0.4$ dex less than that of inactive galaxies at a given bulge 
    luminosity. The zero point offset in the \mlr\ relation of elliptical 
    and classical bulges in our sample compared to that in inactive galaxies 
    appears to be correlated with the Eddington ratio.
    \item For ellipticals and classical bulges, our sample follows the 
    Kormendy relation in a similar manner as normal galaxies, indicating 
    there is no evidence for overluminous bulges in our sample. As a result, 
    we conclude that the zero point offset is possibly due to the dependence 
    of the scaling factor on the Eddington ratio.
    \item Pseudo-bulges in our sample AGNs tend to be overluminous compared 
    to those in normal galaxies, which is inferred from the Kormendy relation. 
    This property of the pseudo-bulges is possibly due to the young stellar populations in the AGN host.   
    
\end{itemize}

\acknowledgments

We thank the anonymous referee for the constructive feedback that helped to improve the quality of the paper.
This work is based on observations made with the NASA/ESA Hubble Space 
Telescope, obtained at the Space Telescope Science Institute, which is 
operated by the Association of Universities for Research in Astronomy, Inc., 
under NASA contract NAS5-26555. These observations are associated with 
program \#15444.
LCH was supported by the National Science Foundation of China (11721303, 
11991052, 12011540375).), China Manned Space Project (CMS-CSST-2021-A04), 
and the National Key R\&D Program of China (2016YFA0400702). This work was 
supported by a National Research Foundation of 
Korea (NRF) grant (No.\ 2020R1A2C4001753) funded by the Korean government 
(MSIT) and under the framework of international cooperation program managed 
by the National Research Foundation of Korea (NRF-2020K2A9A2A06026245).  
This research made use of the ``{\it k-}corrections calculator'' service 
available at http://kcor.sai.msu.ru/.

\begin{table*}
\centering
\caption{
The Sample.
\label{tab:table1}}
\begin{tabular}{llcccccc}
\toprule
\ct{Source Name} & \ct{Alternative Name} & Exposure & \ct{R.A.} & \ct{Dec.} & $A_{\rm F814W}$ & $z$ &  \ct{$D_{\rm L}$} \\
 & & (s) & \ct{(deg.)} & \ct{(deg.)} & \ct{(mag)} & & \ct{(Mpc)} \\ 
\ct{(1)} & \ct{(2)} & \ct{(3)} & \ct{(4)} & \ct{(5)} & \ct{(6)} & \ct{(7)} & \ct{(8)} \\
\midrule
SWIFT J0123.9$-$5846 & Fairall 9 & 674 & $\;\:20.9408$ & $-58.8057$ & 0.039 & 0.0460 & 210.4 \\
SWIFT J0157.2+4715 & 2MASX J01571097+4715588 & 674 & $\;\:29.2956$ & $\;\;\:47.2666$ & 0.228 & 0.0478 & 218.9 \\
SWIFT J0206.2$-$0019 & Mrk 1018 & 674 & $\;\:31.5666$ & $\;\:-0.2914$ & 0.042 & 0.0430 & 196.2 \\
SWIFT J0234.6$-$0848 & NGC 985 & 674 & $\;\:38.6574$ & $\;\:-8.7876$ & 0.050 & 0.0430 & 196.4 \\
SWIFT J0333.3+3720 & 2MASX J03331873+3718107 & 674 & $\;\:53.3282$ & $\;\;\:37.3030$ & 0.815 & 0.0547 & 251.8 \\
SWIFT J0429.6$-$2114 & 2MASX J04293830$-$2109441 & 674 & $\;\:67.4095$ & $-21.1622$ & 0.037 & 0.0700 & 325.8 \\
SWIFT J0510.7+1629 & IRAS 05078+1626 & 674 & $\;\:77.6896$ & $\;\;\:16.4989$ & 0.457 & 0.0173 & $\;\:77.7$ \\
SWIFT J0516.2$-$0009 & Ark 120 & 674 & $\;\:79.0476$ & $\;\:-0.1498$ & 0.194 & 0.0325 & 147.1 \\
SWIFT J0736.9+5846 & Mrk 9 & 674 & 114.2374 & $\;\;\:58.7704$ & 0.089 & 0.0398 & 181.5 \\
SWIFT J0743.3$-$2546 & LEDA 86073 & 674 & 115.8114 & $-25.7639$ & 1.096 & 0.0238 & 106.5 \\
SWIFT J0747.5+6057 & Mrk 10 & 554 & 116.8714 & $\;\;\:60.9335$ & 0.071 & 0.0294 & 131.6 \\
SWIFT J0759.8$-$3844 & 2MASS J07594181$-$3843560 & 674 & 119.9242 & $-38.7322$ & 1.238 & 0.0402 & 183.2 \\
SWIFT J0923.7+2255 & MCG +04$-$22$-$042 & 674 & 140.9292 & $\;\;\:22.9090$ & 0.067 & 0.0333 & 150.8 \\
SWIFT J0942.2+2344 & CGCG 122$-$055 & 674 & 145.5200 & $\;\;\:23.6853$ & 0.038 & 0.0217 & $\;\:97.4$ \\
SWIFT J1020.5$-$0237B & 2MASX J10195855$-$0234363 & 674 & 154.9941 & $\;\:-2.5767$ & 0.062 & 0.0595 & 274.7 \\
SWIFT J1132.9+1019A & IC 2921 & 674 & 173.2053 & $\;\;\:10.2965$ & 0.050 & 0.0440 & 201.0 \\
SWIFT J1139.1+5913 & SBS 1136+594 & 674 & 174.7870 & $\;\;\:59.1990$ & 0.023 & 0.0616 & 284.8 \\
SWIFT J1143.7+7942 & UGC 06728 & 674 & 176.3168 & $\;\;\:79.6815$ & 0.155 & 0.0063 & $\;\:28.1$ \\
SWIFT J1148.3+0901 & 2MASX J11475508+0902284 & 674 & 176.9795 & $\;\;\:\;\:9.0413$ & 0.041 & 0.0693 & 322.2 \\
SWIFT J1316.9$-$7155 & 2MASX J13165424$-$7155270 & 674 & 199.2262 & $-71.9242$ & 0.389 & 0.0703 & 327.2 \\
SWIFT J1349.7+0209 & UM 614 & 674 & 207.4701 & $\;\;\:\;\:2.0791$ & 0.043 & 0.0331 & 150.0 \\
SWIFT J1416.9$-$1158 & 2MASX J14165001$-$1158577 & 674 & 214.2084 & $-11.9829$ & 0.102 & 0.0992 & 471.1 \\
SWIFT J1421.4+4747 & SBS 1419+480 & 674 & 215.3742 & $\;\;\:47.7902$ & 0.027 & 0.0727 & 338.8 \\
SWIFT J1747.7$-$2253 & 2MASS J17472972$-$2252448 & 674 & 266.8739 & $-22.8791$ & 1.555 & 0.0467 & 213.9 \\
SWIFT J1747.8+6837A & Mrk 507 & 674 & 267.1599 & $\;\;\:68.7044$ & 0.059 & 0.0551 & 253.9 \\
SWIFT J1747.8+6837B & VII Zw 742 & 674 & 266.7493 & $\;\;\:68.6102$ & 0.057 & 0.0630 & 291.5 \\
SWIFT J1844.5$-$6221 & Fairall 51 & 674 & 281.2249 & $-62.3648$ & 0.165 & 0.0140 & $\;\:62.3$ \\
SWIFT J2035.2+2604 & 2MASX J20350566+2603301 & 674 & 308.7735 & $\;\;\:26.0583$ & 0.415 & 0.0478 & 218.9 \\
SWIFT J2044.0+2832 & RX J2044.0+2833 & 674 & 311.0188 & $\;\;\:28.5534$ & 0.524 & 0.0489 & 224.0 \\
SWIFT J2109.1$-$0942 & 2MASX J21090996$-$0940147 & 674 & 317.2915 & $\;\:-9.6707$ & 0.325 & 0.0267 & 120.7 \\
SWIFT J2114.4+8206 & 2MASX J21140128+8204483 & 674 & 318.5049 & $\;\;\:82.0801$ & 0.233 & 0.0833 & 391.0 \\
SWIFT J2118.9+3336 & 2MASX J21192912+3332566 & 674 & 319.8714 & $\;\;\:33.5491$ & 0.328 & 0.0509 & 233.7 \\
SWIFT J2124.6+5057 & 4C 50.55 & 674 & 321.1643 & $\;\;\:50.9735$ & 3.709 & 0.0151 & $\;\:67.4$ \\
SWIFT J2156.1+4728 & 2MASX J21355399+4728217 & 674 & 323.9750 & $\;\;\:47.4727$ & 0.969 & 0.0253 & 114.0 \\
SWIFT J2219.7+2614 & 2MASX J22194971+2613277 & 674 & 334.9573 & $\;\;\:26.2244$ & 0.165 & 0.0877 & 413.2 \\
\end{tabular}
\tabnote{
Col. (1): Source name.
Col. (2): Alternative name.
Col. (3): Exposure time.
Col. (4): Right Ascension.
Col. (5): Declination.
Col. (6): Galactic extinction in F814W.
Col. (7): Redshift.
Col. (8): Luminosity distance.
}
\end{table*}

\begin{table*}
\centering
\caption{
Physical Properties of the Sample.
\label{tab:table2}}
\begin{tabular}{lccccc}
\toprule
\ct{Source Name} & Line & Bulge Type & $\log M_{\rm BH}$ & $\log L_{\rm bol}$ & \ct{$\log \lambda_{\rm E}$} \\
 & & & ($M_\odot$) & (erg $\rm s^{-1}$) & \\ 
\ct{(1)} & (2) & (3) & (4) & (5) & (6) \\
\midrule
SWIFT J0123.9$-$5846 & \ha & PB & 8.07 & 45.32 & $-0.85$ \\
SWIFT J0157.2+4715   & \ha & PB & 7.54 & 44.84 & $-0.80$ \\
SWIFT J0206.2$-$0019 & \ha & E & 8.11 & 45.01 & $-1.20$ \\
SWIFT J0234.6$-$0848 & \ha & PB & 8.13 & 45.02 & $-1.21$ \\
SWIFT J0333.3+3720   & \ha & E & 8.25 & 45.14 & $-1.21$ \\
SWIFT J0429.6$-$2114 & \ha & E & 8.83 & 45.00 & $-1.93$ \\
SWIFT J0510.7+1629   & \ha & E & 7.76 & 44.72 & $-1.14$ \\
SWIFT J0516.2$-$0009 & \ha & CB & 8.78 & 45.11 & $-1.77$ \\
SWIFT J0736.9+5846   & \ha & PB & 7.53 & 44.40 & $-1.23$ \\
SWIFT J0743.3$-$2546 & \ha & PB & 7.09 & 44.30 & $-0.89$ \\
SWIFT J0747.5+6057   & \ha & PB & 7.24 & 44.36 & $-0.98$ \\
SWIFT J0759.8$-$3844 & \ha & E & 8.62 & 45.15 & $-1.57$ \\
SWIFT J0923.7+2255   & \ha & PB & 7.23 & 44.87 & $-0.46$ \\
SWIFT J0942.2+2344   & \ha & PB & 7.04 & 43.98 & $-1.16$ \\
SWIFT J1020.5$-$0237B& \hb & PB & 8.29 & 44.60 & $-1.79$ \\
SWIFT J1132.9+1019A  & \ha & CB & 7.95 & 44.72 & $-1.33$ \\
SWIFT J1139.1+5913   & \ha & E & 8.31 & 45.14 & $-1.27$ \\
SWIFT J1143.7+7942   & \ha & PB & 5.71 & 43.28 & $-0.53$ \\
SWIFT J1148.3+0901   & \ha & PB & 8.33 & 45.02 & $-1.41$ \\
SWIFT J1316.9$-$7155 & \hb & E & 9.15 & 45.16 & $-2.09$ \\
SWIFT J1349.7+0209   & \ha & CB & 7.61 & 44.51 & $-1.20$ \\
SWIFT J1416.9$-$1158 & \hb & E & 9.17 & 45.53 & $-1.74$ \\
SWIFT J1421.4+4747   & \ha & CB & 8.65 & 45.27 & $-1.48$ \\
SWIFT J1747.7$-$2253 & \ha & E & 9.09 & 44.93 & $-2.26$ \\
SWIFT J1747.8+6837A  & \ha & PB & 6.91 & 44.35 & $-0.66$ \\
SWIFT J1747.8+6837B  & \ha & PB & 6.90 & 44.59 & $-0.41$ \\
SWIFT J1844.5$-$6221 & \ha & PB & 6.90 & 44.15 & $-0.85$ \\
SWIFT J2035.2+2604   & \ha & CB & 7.63 & 44.67 & $-1.06$ \\
SWIFT J2044.0+2832   & \ha & PB & 7.93 & 44.93 & $-1.10$ \\
SWIFT J2109.1$-$0942 & \ha & PB & 7.22 & 44.40 & $-0.92$ \\
SWIFT J2114.4+8206   & \ha & E & 9.13 & 45.68 & $-1.55$ \\
SWIFT J2118.9+3336   & \ha & E & 8.22 & 44.77 & $-1.55$ \\
SWIFT J2124.6+5057   & \ha & E & 6.80 & 44.95 & $\;\;\:0.05$ \\
SWIFT J2156.1+4728   & \ha & CB & 7.58 & 44.39 & $-1.29$ \\
SWIFT J2219.7+2614   & \ha & E & 9.12 & 45.41 & $-1.81$ \\
\end{tabular}
\tabnote{
Col. (1): Source name.
Col. (2): Line used to estimate BH mass.
Col. (3): Bulge type: "E"=elliptical, "CB"=classical bulge, "PB"=pseudo-bulge.
Col. (4): BH mass.
Col. (5): Bolometric luminosity inferred from the intrinsic X-ray luminosity (\citealt{ricci_2017}).
Col. (6): Eddington ratio.
}
\end{table*}

\begin{table*}
\centering
\caption{
Photometric Properties of the Sample. 
\label{tab:table3}}
\begin{tabular}{lcccccccc}
\toprule
 & Nuclear & \multicolumn{4}{c}{Bulge} & \ct{Host} & & \\ [0.5ex] 
\cline{3-6} 
 \addlinespace[0.5ex] 
 \ct{Source Name} & \ct{$M_I$} & \ct{$M_I$} & \ct{$n$} & \ct{$R_{\rm e}$} & \ct{$\langle\mu_{\rm e}\rangle$} & \ct{$M_I$} & \ct{$L_{\rm bul}$/$L_{\rm nuc}$} & \ct{$B/T$} \\
& \ct{(mag)} & \ct{(mag)} & & \ct{(arcsec)} & \ct{(mag arcsec$^{-2}$)} & \ct{(mag)} & & \\
\ct{(1)} & \ct{(2)} & \ct{(3)} & \ct{(4)} & \ct{(5)} & \ct{(6)} & \ct{(7)} & \ct{(8)} & \ct{(9)} \\
\midrule
SWIFT J0123.9$-$5846 & $-21.68$ & $-22.00\pm0.40$ & 0.99 &  $\;\:0.61$ & 15.50 & $-23.77$ & \quad1.34 & 0.20 \\
SWIFT J0157.2+4715   & $-19.42$ & $-21.13\pm0.40$ & 1.47 &  $\;\:0.75$ & 16.80 & $-22.89$ & \quad4.83 & 0.20 \\
SWIFT J0206.2$-$0019 & $-19.53$ & $-23.90\pm0.30$ & 4(f) &  $\;\:8.58$ & 19.33 & $-23.90$ & $\;\:55.94$ & 1.00 \\
SWIFT J0234.6$-$0848 & $-21.70$ & $-21.78\pm0.40$ & 1.48 &  $\;\:1.75$ & 18.00 & $-23.56$ & \quad1.08 & 0.19 \\
SWIFT J0333.3+3720   & $-21.68$ & $-23.02\pm0.30$ & 4(f) &  $\;\:7.98$ & 20.05 & $-23.02$ & \quad3.45 & 1.00 \\
SWIFT J0429.6$-$2114 & $-21.72$ & $-23.62\pm0.30$ & 4(f) &  $\;\:9.30$ & 19.79 & $-23.62$ & \quad5.71 & 1.00 \\
SWIFT J0510.7+1629   & $-19.54$ & $-20.96\pm0.30$ & 4(f) &  $\;\:3.10$ & 20.06 & $-20.96$ & \quad3.70 & 1.00 \\
SWIFT J0516.2$-$0009 & $-22.21$ & $-22.96\pm0.40$ & 4(f) &  $\;\:2.65$ & 17.72 & $-23.32$ & \quad1.99 & 0.72 \\
SWIFT J0736.9+5846   & $-21.01$ & $-19.99\pm0.40$ & 0.70 &  $\;\:0.38$ & 16.50 & $-22.86$ & \quad0.39 & 0.07 \\
SWIFT J0743.3$-$2546 & $-19.34$ & $-20.11\pm0.40$ & 2.17 &  $\;\:0.59$ & 17.30 & $-22.54$ & \quad2.04 & 0.11 \\
SWIFT J0747.5+6057   & $-18.67$ & $-20.77\pm0.40$ & 1.50 &  $\;\:0.82$ & 17.36 & $-23.28$ & \quad6.89 & 0.10 \\
SWIFT J0759.8$-$3844 & $-21.43$ & $-21.83\pm0.30$ & 4(f) &  $\;\:1.88$ & 18.11 & $-21.83$ & \quad1.44 & 1.00 \\
SWIFT J0923.7+2255   & $-19.45$ & $-20.91\pm0.40$ & 1.37 &  $\;\:0.66$ & 16.76 & $-22.77$ & \quad3.85 & 0.18 \\
SWIFT J0942.2+2344   & $-18.45$ & $-19.74\pm0.40$ & 0.96 &  $\;\:0.30$ & 16.19 & $-21.51$ & \quad3.28 & 0.20 \\
SWIFT J1020.5$-$0237B& $-19.76$ & $-21.13\pm0.40$ & 1.31 &  $\;\:0.94$ & 17.30 & $-23.29$ & \quad3.52 & 0.14 \\
SWIFT J1132.9+1019A  & $-18.77$ & $-21.37\pm0.40$ & 2.66 &  $\;\:0.97$ & 17.14 & $-22.47$ & $\;\:10.99$ & 0.36 \\
SWIFT J1139.1+5913   & $-21.69$ & $-20.90\pm0.30$ & 4(f) &  $\;\:2.45$ & 19.61 & $-20.90$ & \quad0.48 & 1.00 \\
SWIFT J1143.7+7942   & $-16.62$ & $-18.42\pm0.40$ & 1.35 &  $\;\:0.30$ & 17.54 & $-20.35$ & \quad5.21 & 0.17 \\
SWIFT J1148.3+0901   & $-20.36$ & $-20.97\pm0.40$ & 1.39 &  $\;\:0.50$ & 16.10 & $-22.37$ & \quad1.76 & 0.27 \\
SWIFT J1316.9$-$7155 & $-21.98$ & $-23.19\pm0.30$ & 4(f) &  $\;\:4.17$ & 18.48 & $-23.19$ & \quad3.05 & 1.00 \\
SWIFT J1349.7+0209   & $-19.08$ & $-20.53\pm0.40$ & 4(f) &  $\;\:0.57$ & 16.82 & $-21.68$ & \quad3.81 & 0.35 \\
SWIFT J1416.9$-$1158 & $-22.94$ & $-23.69\pm0.30$ & 4(f) &  $\;\:7.31$ & 19.19 & $-23.69$ & \quad2.01 & 1.00 \\
SWIFT J1421.4+4747   & $-22.34$ & $-21.89\pm0.40$ & 2.04 &  $\;\:1.01$ & 16.68 & $-22.75$ & \quad0.66 & 0.46 \\
SWIFT J1747.7$-$2253 & $-20.84$ & $-22.72\pm0.30$ & 4(f) &  $\;\:5.38$ & 19.50 & $-22.72$ & \quad5.67 & 1.00 \\
SWIFT J1747.8+6837A  & $-19.95$ & $-20.21\pm0.40$ & 1.21 &  $\;\:0.25$ & 15.31 & $-22.47$ & \quad1.27 & 0.13 \\
SWIFT J1747.8+6837B  & $-21.71$ & $-21.63\pm0.40$ & 2.66 &  $\;\:2.55$ & 18.97 & $-23.87$ & \quad0.92 & 0.13 \\
SWIFT J1844.5$-$6221 & $-18.99$ & $-19.71\pm0.40$ & 1.23 &  $\;\:0.37$ & 16.70 & $-21.74$ & \quad1.95 & 0.15 \\
SWIFT J2035.2+2604   & $-19.56$ & $-21.78\pm0.40$ & 3.24 &  $\;\:2.18$ & 18.48 & $-22.73$ & \quad7.69 & 0.42 \\
SWIFT J2044.0+2832   & $-21.95$ & $-20.81\pm0.40$ & 0.77 &  $\;\:0.88$ & 17.47 & $-22.32$ & \quad0.35 & 0.25 \\
SWIFT J2109.1$-$0942 & $-20.69$ & $-21.22\pm0.40$ & 1.47 &  $\;\:0.64$ & 16.37 & $-22.25$ & \quad1.63 & 0.39 \\
SWIFT J2114.4+8206   & $-23.23$ & $-23.57\pm0.30$ & 4(f) & 10.73 & 20.15 & $-23.57$ & \quad1.37 & 1.00 \\
SWIFT J2118.9+3336   & $-18.40$ & $-23.67\pm0.30$ & 4(f) & 10.33 & 19.97 & $-23.67$ & 127.82    & 1.00 \\
SWIFT J2124.6+5057   & $-20.96$ & $-21.33\pm0.30$ & 4(f) &  $\;\:0.99$ & 17.22 & $-21.33$ & \quad1.41 & 1.00 \\
SWIFT J2156.1+4728   & $-19.26$ & $-20.29\pm0.40$ & 2.73 &  $\;\:1.11$ & 18.50 & $-21.73$ & \quad2.57 & 0.27 \\
SWIFT J2219.7+2614   & $-21.45$ & $-24.07\pm0.30$ & 4(f) & 14.93 & 20.36 & $-24.07$ & $\;\:11.16$ & 1.00 \\
\end{tabular}
\tabnote{
Col. (1): Source name.
Col. (2): Absolute $I$-band luminosity of nucleus.
Col. (3): Absolute $I$-band luminosity of bulge.
Col. (4): \ser\ index of bulge; ``(f)'' implies that $n$ is fixed to 4.
Col. (5): Effective radius of bulge.
Col. (6): Mean surface brightness within the effective radius.
Col. (7): Absolute $I$-band luminosity of host galaxy.
Col. (8): The luminosity ratio of bulge to nucleus.
Col. (9): Bulge-to-total light ratio.
}
\end{table*}

\begin{table*}
\centering
\caption{
\mlr\ Relation for Various Subsamples.
\label{tab:table4}}
\begin{tabular}{lccc}
\toprule
\ct{Subsamples} & $\alpha$ & $\beta$ & $\epsilon_0$ \\
\ct{(1)} & (2) & (3) & (4) \\
\midrule
Inactive (All) & $-0.57$ & $-4.17\pm0.82$ & $0.53\pm0.06$\\
AGNs (All) & $-0.57$ & $-4.40 \pm 0.09$ & $0.23 \pm 0.10$ \\
AGNs (All; $\lambda_{\rm E} \leq 0.06$) & $-0.57$ & $-4.19 \pm 0.10$ & $0.11 \pm 0.11$ \\
AGNs (All; $\lambda_{\rm E} > 0.06$) & $-0.57$ & $-4.64 \pm 0.10$ & $0.12 \pm 0.09$ \\
 & & & \\
Inactive (E+CB) & $-0.49$ & $-2.22\pm0.62$ & $0.31\pm0.03$\\
AGNs (E+CB) & $-0.49$ & $-2.62 \pm 0.13$ & $0.28 \pm 0.12$ \\
AGNs (E+CB; $\lambda_{\rm E} \leq 0.06$) & $-0.49$ & $-2.46 \pm 0.11$ & $0.03 \pm 0.09$ \\
AGNs (E+CB; $\lambda_{\rm E} > 0.06$) & $-0.49$ & $-3.06 \pm 0.23$ & $0.38 \pm 0.14$ \\
 & & & \\
Inactive (PB) & $-0.49$ & $-2.92\pm0.14$ & $0.62\pm0.14$ \\
AGNs (PB) & $-0.49$ & $-2.74 \pm 0.11$ & $0.13 \pm 0.11$ \\
AGNs (PB; $\lambda_{\rm E} \leq 0.06$) & $-0.49$ & $-2.20 \pm 0.10$ & $0.00 \pm 0.00$ \\
AGNs (PB; $\lambda_{\rm E} > 0.06$) & $-0.49$ & $-2.92 \pm 0.10$ & $0.00 \pm 0.06$ \\
\end{tabular}
\tabnote{
Col. (1): Subsample.
Col. (2): Slope.
Col. (3): Zero point.
Col. (4): Intrinsic scatter.
}
\end{table*}

\begin{table*}
\centering
\caption{
Kormendy Relation for Various Subsamples.
\label{tab:table5}}
\begin{tabular}{lcc}
\toprule
\ct{Subsamples} & $\kappa$ & $\gamma$ \\
\ct{(1)} & (2) & (3) \\
\midrule
Inactive (E) [G] & $2.38 \pm 0.07$ & $17.16$ \\
Inactive (E+CB) & $2.53 \pm 0.06$ & $17.66$ \\
Inactive (PB) & $3.94 \pm 0.29$ & $18.94$ \\
 & & \\
AGNs (All) & $2.56 \pm 0.20$ & $17.44$ \\
AGNs (E+CB) & $2.47 \pm 0.33$ & $17.51$ \\
AGNs (E+CB; $\lambda_{\rm E} \leq 0.06$) & $2.51 \pm 0.40$ & $17.43$ \\
AGNs (E+CB; $\lambda_{\rm E} > 0.06$) & $2.57 \pm 0.91$ & $17.60$ \\
 & & \\
AGNs (PB) & $2.52 \pm 0.61$ & $17.40$ \\
AGNs (PB; $\lambda_{\rm E} \leq 0.06$) & $2.67 \pm 0.71$ & $17.31$ \\
AGNs (PB; $\lambda_{\rm E} > 0.06$) & $2.53 \pm 0.81$ & $17.44$ \\
\end{tabular}
\tabnote{
Col. (1): Subsample; ``[G]'' denotes that the subsample comes from \citet{gao_2020}.
Col. (2): Slope.
Col. (3): Zero point.
}

\end{table*}





\end{document}